\documentclass[12pt, draftclsnofoot, onecolumn]{IEEEtran}
\usepackage{graphicx}
\usepackage{epstopdf}
\usepackage{ifpdf}
\usepackage{subfig}
\usepackage[font=scriptsize]{caption}
\usepackage{amssymb}
\usepackage{amsmath}
\usepackage{mathrsfs}
\usepackage{longtable}
\usepackage{tabularx}
\usepackage{blindtext}
\usepackage{cite}
\usepackage{color}
\usepackage{url}
\usepackage{algorithm}
\usepackage{algorithmicx, algpseudocode}
\usepackage{balance}

\newcommand{\la}{\langle}
\newcommand{\ra}{\rangle}

\DeclareMathOperator{\tr}{\mathrm{Tr}}
\DeclareMathOperator{\loga}{\mathrm{log}}

\newcommand{\ds}{\displaystyle}
\newcommand{\st}{\mbox{s.t.}}

\newcommand{\Hc}[2]{\pmb{h}_{{#1}{#2}}}

\newcommand{\Fp}[2]{\pmb{f}_{{#1}{#2}}}

\newcommand{\nn}[1]{n_{#1}}

\newcommand{\clK}{{\cal K}}
\newcommand{\clI}{{\cal I}}
\newcommand{\bF}{\mathbf{F}}
\newcommand{\bp}{\mathbf{p}}
\newcommand{\bfs}{\mathbf{s}}
\newcommand{\bx}{\mathbf{x}}
\newcommand{\bH}{\mathbf{H}}
\newcommand{\bh}{\mathbf{h}}
\newcommand{\bfy}{\mathbf{y}}
\newcommand{\bn}{\mathbf{n}}
\newcommand{\bs}{\mathbf{s}}
\newcommand{\bff}{\mathbf{f}}

\newcommand{\Col}{{\sf Col}}
\newcommand{\bX}{\mathbf{X}}
\newcommand{\bA}{\mathbf{A}}
\newcommand{\Row}{{\sf Row}}
\begin{document}
\title{Multi-cell Massive MIMO Beamforming in Assuring QoS for Large Numbers of Users 
\thanks{This work was supported in part by the U.K. Royal Academy of Engineering Research Fellowship under Grant RF1415/14/22 and U.K. Engineering and Physical Sciences Research Council under Grant EP/P019374/1. The work of H. D. Tuan was supported in part by the Australian Research Councils Discovery Projects under Project DP130104617. The work of H. V. Poor was supported in part by the U.S. National Science Foundation under Grants CNS-1456793 and ECCS-1343210}}
\author{L. D. Nguyen,
\thanks{Long D. Nguyen and Trung Q. Duong are with
Queen's University, Belfast BT7 1NN, UK (e-mail:\{lnguyen04,trung.q.duong\}@qub.ac.uk)} H. D. Tuan\thanks{
Hoang D. Tuan is with the School of Electrical and Data Engineering, University of Technology Sydney, Sydney, NSW 2007, Australia (e-mail: Tuan.Hoang@uts.edu.au)}, T. Q. Duong and H. V. Poor\thanks{H. Vincent Poor is with the Department of Electrical Engineering, Princeton University, Princeton, NJ 08544, USA (e-mail: poor@princeton.edu) }}
\maketitle
\vspace*{-1.5cm}
\begin{abstract}
Massive multi-input multi-output (MIMO) uses a very large number of low-power transmit antennas to serve much smaller numbers of users. The most widely proposed type of massive MIMO transmit beamforming is zero-forcing, which is based on the right inverse of the overall MIMO channel matrix to force the inter-user interference to zero. The performance of massive MIMO is then analyzed based on the throughput of cell-edge users. This paper reassesses  this beamforming philosophy, to instead consider the maximization of 
 the energy efficiency of massive MIMO systems in assuring the quality-of-service (QoS) for as many
users as possible. The bottleneck of serving small numbers of users by a large number of transmit antennas is unblocked
by a new time-fraction-wise beamforming technique, which focuses signal transmission  in fractions of a time slot.
Accordingly, massive MIMO can deliver better  quality-of-experience (QoE) in assuring QoS for much larger numbers of users. The provided
simulations show that the numbers of users served by massive MIMO with the required QoS may be twice or more than
the number of its transmit antennas.
\end{abstract}

\section{Introduction}
Massive multi-input multi-output (MIMO) \cite{rusek2013, Larsson2014} is a potential next-generation communication technology, which can promise quality-of-service (QoS) for cell edge users. As envisioned in the pioneering work \cite{M10}, massive MIMO is meant to serve smaller numbers of users by a large array of low-power transmit antennas.
Under such an environment, massive MIMO exhibits favorable propagation characteristics, i.e., orthogonality of communication channels \cite{rusek2013,HLM14} and  deterministic behavior of the channels' eigenvalue distribution \cite{TV04,CD11},
which allow low-complexity zero-forcing (ZF) beamforming to perform well \cite{Waetal12,YM13}. The performance analysis of
 such ZF beamforming is typically based on the equi-power allocation among beamformers \cite{Lim2015}. Our recent work \cite{NTDP17} shows that the users' QoS can increase significantly
 by employing the optimal power allocation among beamformers. Equally importantly, it also shows that the optimal 
 power-allocated ZF beamforming performs much better than  optimal power-allocated
conjugate beamforming though the latter seems to perform better than the former under the equi-power allocation \cite{M10,YM13}. To serve many users, massive MIMO must schedule its service. Thus, small numbers of users are served at 
any given time. As such, it is not known if massive MIMO is able to deliver a quality-of-experience to many users simultaneously.

The involvement of more users results in  ill-conditioning of the right-inverse of the channel matrices, which can be overcome by the so called regularized zero-forcing (RZF) beamforming \cite{PHS05,Waetal12}. However, by employing RZF, the inter-user interference can no longer be forced to zero and its impact on the performance of RZF beamforming must be addressed. Another issue with massive MIMO is that its transmit antennas, which are closely packed in a very small space, are sometimes 
assumed to be spatially uncorrelated. Under this assumption, the channel matrices are well-conditioned and the zero-forcing beamformers are expected to perform well according to the power-scaling law \cite{Hien2013Jou}. However, due to the scattering environment, these antennas are inherently spatially correlated \cite{Shetal00,Adhi_TIT_2013},
lowering the rank of the channel matrices and thus affecting the capacity of massive MIMO.

In this paper we consider the problem of maximizing the massive MIMO's energy efficiency (EE) under users' QoS constraints (in terms of their throughput thresholds) and a transmit power budget, which is motivated by the following concerns:
\begin{itemize}
\item The EE  in terms of the ratio between the total information throughput and the total consumed power is an important metric for assessing the performance of futuristic communication systems \cite{Buetal16,Zaetal16}.
\item In massive MIMO systems, the EE is particularly important to control the scale of the antenna arrays, which should
 generally to be as large as possible to gain more benefits from the transmit power-scaling law \cite{Hien2013Jou}. Larger scaled arrays consume more circuit power, which is linearly proportional to the number of their antennas. Reducing circuit power consumed by hardware requires the reduction of radio frequency chains which not only leads to a  complicated signal transmission but also
makes  ZF beamforming for massive MIMO  lose both its simplicity in design and efficiency in information delivery. More importantly, the multi-channel diversity of massive MIMO is limited by the number
of radio frequency chains used.
\item Addressing the EE under users' QoS constraints achieves simultaneous optimization for power and network throughput in assuring users' QoS. It is important to emphasize here that the capacity of massive MIMO in serving many users considered in this paper is different from \cite{YG06}, which considers the system sum throughput without users' QoS and as such most of the throughput would be enjoyed by a few users with stronger channels.
\end{itemize}
Our contributions are  as follows:
\begin{itemize}
\item We develop new path-following algorithms for computation of the EE maximization problem subject to users' QoS constraints under practical scenarios of massive MIMO, where the antennas' spatial correlation is incorporated;
\item To assure QoS for as many users as possible, we propose a time-fraction-wise transmit beamforming scheme, which assures the QoS for  users within fractions of a time slot.
    This novel beamforming scheme
relies on a much more complex optimization problem. Nevertheless, we develop a new path-following algorithm tailored
for its computation. Our simulation shows that massive MIMO equipped with large antenna arrays is able to assure
the QoS for even much larger numbers of users.
\end{itemize}
The paper is organized as follows. ZF and RZF beamforming to assure the users' QoS is considered in Section II.
Section III is devoted to time-fraction-wise ZF and RZF beamforming.
Simulations are provided in Section IV and conclusions are given in Section V.
Appendix provides some important inequalities that are used in the algorithmic developments.

\emph{Notation.} Boldface upper and lowercase letters denote matrices and vectors, respectively. The transpose and conjugate transpose of a matrix $\pmb{X}$ are respectively represented by $\pmb{X}^T$ and $\pmb{X}^H$. $\pmb{I}$ and $\pmb{0}$ stand for identity and zero matrices of appropriate dimensions. $\tr (.)$ is the trace operator. $||\pmb{x}||$ is the Euclidean norm of the vector $\pmb{x}$ and $||\pmb{X}||$ is the Frobenius norm of the matrix $\pmb{X}$.  A Gaussian random vector with mean $\bar{\pmb{x}}$ and covariance $\pmb{R}_{\pmb{x}}$ is denoted by $\pmb{x}\sim \mathcal{CN}(\bar{\pmb{x}},\pmb{R}_{\pmb{x}})$.
For matrices $\pmb{X}_i$, $i=1, \dots, \pmb{X}_k$ of appropriate dimension, $\Col[\bX_i]_{i=1,\dots,K}$ or
$\Col[\bX_i]_{i\in\clK}$ for $\clK\triangleq \{1,\dots, k\}$ arranges $\bX_i$ in block column, i.e.
\[
\Col[\bX_i]_{i\in\clK}=\begin{bmatrix}\pmb{X}_1\cr
 \dots\cr
\pmb{X}_k
\end{bmatrix}
\]
so it is true that $\Col[\bX_i]_{i\in\clK}\bA=\Col[\bX_i\bA]_{i\in\clK}$. Analogously,
$\Row[\bX_i]_{i=1,\dots,K}$ or
$\Row[\bX_i]_{i\in\clK}$  arranges $\bX_i$ in block row, i.e.
\[
\Row[\bX_i]_{i\in\clK}=\begin{bmatrix}\pmb{X}_1&
 \dots&
\pmb{X}_k
\end{bmatrix}
\]
so it is true that $\bA\Row[\bX_i]_{i\in\clK}=\Row[\bA\bX_i]_{i\in\clK}$.
\section{Zero-forcing and regularized zero-forcing beamforming}
\begin{figure}[H]
	\centering
	\centerline{\includegraphics[width=0.7\textwidth]{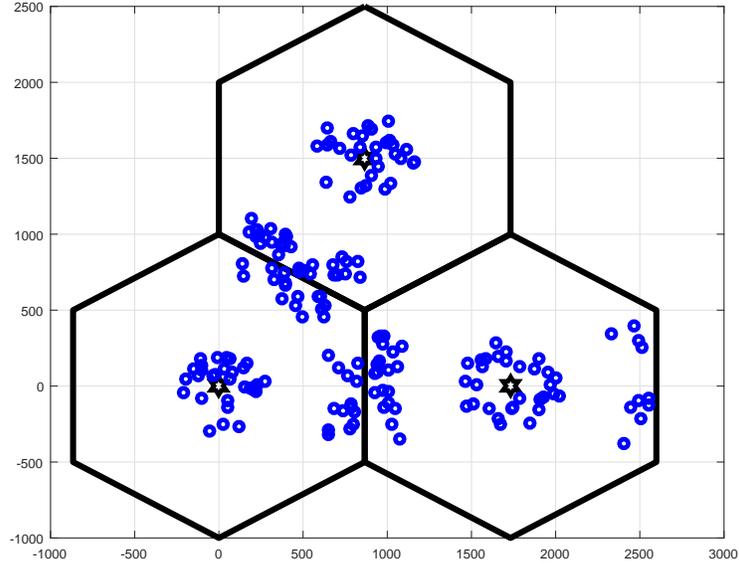}}
	\caption{An equally mixed-coupled three-cell scenario. Each cell has a total of 60 UEs.}
	\label{fig:0_SM_3cell}
\end{figure}

Consider a multi-cell network, which typically consists of three base stations (BSs) as depicted by Fig.
\ref{fig:0_SM_3cell}. Each base station
(BS) $i\in \clI\triangleq \{1, 2, 3\}$ is  equipped with a large-scale $N$ antenna array to serve its $N_{UE}$
single-antenna equipped users (UEs)
$(i,k)$, $k\in \mathcal{K}\triangleq \{1, \dots, N_{UE}\}$ within its cell.
UEs $(i,k)$, $k\in \mathcal{K}_{\sf ne}\triangleq \{1,\dots N_{\sf ne}\}$
are located at a near area to BS $i$ while UE $(i,k)$, $k\in \mathcal{K}_{\sf fa}\triangleq \{N_{\sf ne}+1, \dots, N_{UE}\}$ are located at  cell-edge areas as  Figure \ref{fig:0_SM_3cell} shows. Thus in each cell there are $N_{\sf ne}$ near UEs and
$N_{\sf fa}\triangleq N_{UE}-N_{\sf ne}$ cell-edge UEs.

Denote by $s_{i,k}$ the information from BS $i$ intended for its UE $(i,k)$, which is normalized to $E(|s_{i,k}|^2)=1$.
The vector of information
from BS $i$ intended for all its UEs is defined as $\bfs_i=\Col[s_{i,k}]_{k\in\clK}$. Each $s_{i,k}$ is beamformed
by a vector $\mathbf{f}_{i,k} \in \mathbb{C}^{N}$. The beamforming matrix is defined by
\[
\mathbf{F}_i \triangleq \Row[\mathbf{f}_{i,k}]_{k\in\clK} \in \mathbb{C}^{N \times N_{UE}}.
\]
The signal transmitted from
BS $i$ is $\bx_i=\bF_i\bfs_i$.

The vector channel from BS $j$ to UE $(i,k)$ is modelled by $\sqrt{\beta_{j,i,k}}\Hc{j,i,k}{}$, where
$\sqrt{\beta_{j,i,k}}$ models the path loss and large-scale fading, while
\cite{Adhi_TIT_2013, AdhikaryJSAC_2015, STND17}
\begin{align} \label{channel}
    \mathbf{h}_{j,i,k} = \mathbf{\Theta}_{j}^{1/2} \mathbf{h}_{j,i,k}^w,
\end{align}
where $\mathbf{\Theta}_{j} \in \mathbb{C}^{N \times N}$ is a Hermitian symmetric positive semidefinite spatial correlation matrix of rank $r_{j}$ and $\mathbf{h}_{j,i,k}^w \in \mathbb{C}^{N}$ has independent and identical distributed complex entries of zero mean and unit variance, which represents the small-scale fading.
The channel matrix from BS $j$ to UEs in $i$-th cell is thus
$\beta_{j,i}\bH^H_{j,i}$ where $\beta_{j,i}\triangleq \mbox{diag}[\sqrt{\beta_{j,i,k}}]_{k\in\clK}$ and
\[
\bH^{H}_{j,i}\triangleq \Col[\bh^H_{j,i,k}]_{k\in\clK}.
\]
Let $y_{i,k}\in\mathbb{C}$ be the signal received at UE $(i,k)$ and then $\bfy_i\triangleq \Col[y_{i,k}]_{k\in\clK}$.
The MIMO equation is thus
\begin{eqnarray}
\bfy_i&=&\beta_{i,i}\bH^H_{i,i}\bx_i+\sum_{j\in\clI\setminus\{i\}}\beta_{j,i}\bH^H_{j,i}\bx_j+\bn_i\label{mimo1}\\
&=&\beta_{i,i}\bH^H_{i,i}\bF_i\bs_i+\sum_{j\in\clI\setminus\{i\}}\beta_{j,i}\bH^H_{j,i}\bF_j\bs_j+\bn_i,\label{mimo2}
\end{eqnarray}
where $\bn_i=\Col[n_{i,k}]_{k\in\clK}$ is the noise vector of independent entries $n_{i,k}\in{\cal CN}(0,\sigma^2)$.
Particularly, the  multi-input single output (MISO) equation for the signal received at individual UE $(i,k)$  is
\begin{align}  \label{ue}
     y_{i,k} & = \sqrt{\beta_{i,i,k}}\underbrace{\Hc{i,i,k}{}^H \Fp{i,k}{} s_{i,k}}_{\text{desired signal}}
     + \underbrace{\sum_{\ell \in \mathcal{K}\setminus \{k\}} \sqrt{\beta_{i,i,k}}\Hc{i,i,k}{}^H \Fp{i,\ell}{} s_{i,\ell}}_{\text{inter-user interference}}
     + \underbrace{\sum_{j\in\clI\setminus\{i\}}\sqrt{\beta_{j,i,k}}\bh^H_{j,i,k}\bF_j\bs_j}_{\text{inter-cell interference}}
     + \nn{i,k}.
\end{align}
We seek a beamforming matrix $\bF_i$ in the following class
\begin{equation}\label{base1}
\bF_i=\bar{\bF}_i\mbox{diag}[\sqrt{p_{i,k}}]_{k\in\clK}
\end{equation}
with a predetermined matrix
\begin{equation}\label{base2}
\bar{\bF}_i\triangleq \Row[\bar{\bff}_{i,k}]_{k\in\clK} \in\mathbb{C}^{N\times 2M}.
\end{equation}
For $\bp_i=\Col[p_{i,k}]_{i\in\clK}$ and $\bp=(\bp_i)_{i\in\clI}$,
the inter-user interference and inter-cell interference functions are respectively  defined from (\ref{ue}) as
\begin{equation}\label{IUIs}
\sigma_{i,k}^{\sf U}(\bp_i)
\triangleq \beta_{i,i,k} \sum_{\ell \in \mathcal{K} \setminus \{k\}} |\Hc{i,i,k}{}^H\bar{\bff}_{i,\ell}|^2p_{i,\ell},
\end{equation}
and
\begin{equation}\label{ICIs}
\sigma_{i,k}^{\sf C}(\bp)
\triangleq \beta_{j,i,k} \sum_{j\in\clI\setminus\{i\}} \sum_{\ell \in \mathcal{K}}||\bar{\bff}_{j,\ell}||^2p_{j,\ell}.
\end{equation}
Note that while the intra-cell channel $\Hc{i,i,k}{}$ can be efficiently estimated \cite{STND17},
the intercell-channel $\Hc{j,i,k}{}$ in (\ref{ue}) cannot be estimated and must be defined as in (\ref{ICIs}).
 Under the definitions
\begin{equation}\label{base3}
\alpha_{i,k}\triangleq \beta_{i,i,k} |\Hc{i,i,k}{}^H\bar{\bff}_{i,k}|^2
\end{equation}
and
\begin{equation}\label{base4}
\lambda_{i,k}(\bp)\triangleq \sigma_{i,k}^{\sf U}(\bp_i)+\sigma_{i,k}^{\sf C}(\bp),
\end{equation}
which is a linear function, the information throughput at  UE $(i,k)$ is defined by
\begin{equation} \label{inf}
     r_{i,k}(\bp) = \ln \left( 1 + \frac{\alpha_{i,k} p_{i,k}}{\lambda_{i,k}(\bp) + \sigma^2} \right).
\end{equation}
The transmit power by BS $i$ is the following function, which is also linear in $\bp_i$:
\begin{equation}\label{base5}
\chi_i(\bp_i)=\sum_{k\in\clK}||\bar{\bff}_{i,k}||^2p_{i,k}.
\end{equation}
The entire power consumption for the downlink transmission, which is expressed by
\begin{equation}\label{pbs}
\pi(\bp) = \sum_{i\in\clI}  (\alpha\chi_i(\bp_i)+ N P_{a} + P_{c})
\end{equation}
is an affine function in $\bp$. Here $\alpha > 1$ is the reciprocal of the drain efficiency of the amplifier of BS
 and $P_{a}$ and $P_{c}$ are circuit power per antenna and non-transmission power of the BSs.

The network total throughput is defined as
\[
\varphi(\bp)\triangleq \ds\sum_{(i,k)\in\clI\times \clK}r_{i,k}(\bp).
\]
In this paper, we are interested  in the following EE maximization problem  under QoS constraints and power budgets:
\begin{subequations}\label{opt}
\begin{eqnarray}
\ds\max_{\bp}\ \varphi(\bp)/\pi(\bp)\quad\mbox{s.t.}\label{opta}\\
\chi_i(\bp_i)\leq P_i^{\max}, i\in\clI, \label{optb}\\
r_{i,k}(\bp) \geq \bar{r}_{i,k}, (i,k)\in\clI\times \clK,\label{optc}
\end{eqnarray}
\end{subequations}
where constraints \eqref{optc} set the QoS in terms of the throughput thresholds at each UE and
constraint \eqref{optb} keeps the sum of transmit power  under predefined budgets.

From definition (\ref{inf}) of $r_{i,k}(\bp)$, constraint (\ref{optc})
is equivalent to the linear constraint
\begin{equation} \label{g1ca}
\alpha_{i,k}p_{i,k}\geq (e^{\bar{r}_{i,k}}-1)(\lambda_{i,k}(\bp)+\sigma^2), (i,k)\in\clI\times \clK,
\end{equation}
so (\ref{opt}) is a linear-constrained optimization problem. To obtain a path-following algorithm for solution
of (\ref{opt}), it is most natural to iteratively approximate its objective by a lower bounding concave function
(see e.g. \cite{Shetal17E,STDP17,Ngetal17}).  We now propose a new and simpler approach, which involves
a lower bounding approximation for the function in the numerator of the objective in (\ref{opt}) only but nevertheless also leads to
a path-following computational procedure.

Let $\bp^{(n)}$ be a feasible point for (\ref{opt}) found from the
$(n-1)$th iteration and
\[
t^{(n)}\triangleq \varphi(\bp^{(n)})/\pi(\bp^{(n)}),
\]
so
\begin{equation}\label{new1}
\varphi(\bp^{(n)})-t^{(n)}\pi(\bp^{(n)})=0.
\end{equation}
Using inequality (\ref{ineq4a}) in the Appendix for
\[
x = \alpha_{i,k} p_{i,k},
y =  \lambda_{i,k}(\bp)+\sigma^2,
\]
and
\[
\bar{x}= \alpha_{i,k}p_{i,k}^{(n)},
\bar{y}= \lambda_{i,k}(\bp^{(n)})+\sigma^2,
\]
yields the following lower bounding approximation:
\[
\varphi(\bp)\geq \varphi^{(n)}(\bp)
\]
for
\begin{equation}
\varphi^{(n)}(\bp)\triangleq \ds\sum_{(i,k)\in\clI\times\clK}\left(\bar{a}_{i,k}^{(n)} - \bar{b}_{i,k}^{(n)}/\alpha_{i,k}p_{i,k}  - \bar{c}_{i,k}^{(n)} (\lambda_{i,k}(\bp)+\sigma^2)\right), \label{new1e}
\end{equation}
where
\begin{equation}\label{abcn_new}
\begin{array}{c}
0<\bar{a}_{i,k}^{(n)} \triangleq \ds r_{i,k}(\bp^{(n)}) +
2\alpha_{i,k}p_{i,k}^{(n)}/\left(\lambda_{i,k}(\alpha_{i,k}p_{i,k}^{(n)}+\bp_{i,k}^{(n)})+\sigma^2\right),\\
0<\bar{b}_{i,k}^{(n)} \triangleq \ds (\alpha_{i,k}p_{i,k}^{(n)})^2/\left(\alpha_{i,k}p_{i,k}^{(n)}+\lambda_{i,k}(\bp_{i,k}^{(n)})+\sigma^2\right),\\
0<\bar{c}_{i,k}^{(n)} \triangleq \ds \alpha_{i,k}p_{i,k}^{(n)}/\left(\alpha_{i,k}p_{i,k}^{(n)}+\lambda_{i,k}(\bp_{i,k}^{(n)})+\sigma^2\right)
(\lambda_{i,k}(\bp^{(n)})+\sigma^2).
\end{array}
\end{equation}
At the $n$th iteration, the following convex optimization subproblem is solved to generate the next feasible point $\mathbf{p}^{(n+1)}$ for (\ref{opt}):
\begin{equation} \label{CBRZF_mcell_opt_1}
\ds \max_{\mathbf{p}} \left[ \varphi^{(n)}(\mathbf{p})-t^{(n)}\pi(\bp)\right] \quad
\st  \quad (\ref{optb}), (\ref{g1ca}).
\end{equation}
Note that $\bp^{(n)}$ is a feasible point for (\ref{CBRZF_mcell_opt_1}) satisfying (\ref{new1}). Therefore, as
far as $\bp^{(n+1)}\neq \bp^{(n)}$ we have
\[
\begin{array}{lll}
\varphi^{(n)}(\mathbf{p}^{(n+1)})-t^{(n)}\pi(\bp^{(n+1)})& >&\varphi^{(n)}(\mathbf{p}^{(n)})-t^{(n)}\pi(\bp^{(n)})\\
&=&\varphi(\mathbf{p}^{(n)})-t^{(n)}\pi(\bp^{(n)})\\
&=&0,
\end{array}
\]
which implies
\begin{equation}\label{new2}
t^{(n+1)}\triangleq \varphi^{(n)}(\mathbf{p}^{(n+1)})/\pi(\bp^{(n+1)})>t^{(n)},
\end{equation}
i.e. $\bp^{(n+1)}$ is a better feasible point than $\bp^{(n)}$ for (\ref{opt}). Similarly to \cite[Prop.1]{MW78}
it can be easily shown that at least, Algorithm \ref{alg1} converges to a locally optimal solution of (\ref{d6_rzf_3cell}) satisfying the KKT conditions of optimality.
\begin{algorithm}
	\caption{: Path-following algorithm for solving problem (\ref{opt})} \label{alg1}
	\begin{algorithmic}[1]
		\State \textbf{Initialization}: Solve the following convex optimization problem
\begin{equation}\label{ani1}
\min_{\bp}\max_{i\in\clI}[\chi_i(\bp_i)/P_i^{\max}]\quad\mbox{s.t.}\quad (\ref{g1ca}).
\end{equation}
Exit if its optimal value is more than $1$ because it means that  problem (\ref{opt}) is infeasible.
Otherwise, take its optimal solution as a feasible point $\mathbf{p}^{(0)}$ for the convex constraints (\ref{optb}), (\ref{g1ca}) and set $n:=0$ and $t^{(0)}=\varphi(\bp^{(0)})/\pi(\bp^{(0)})$.
		\State \textbf{Repeat}
		\State Solve  problem (\ref{CBRZF_mcell_opt_1}) for its optimal solution $\mathbf{p}^{(n+1)}$. Set $t^{(n+1)}=\varphi(\bp^{(n+1)})/\pi(\bp^{(n+1)})$.
		\State Set $n:=n+1$.
		\State \textbf{Until} convergence of the objective in (\ref{opt}).
	\end{algorithmic}
\end{algorithm}

\subsection{Zero-forcing and regularized zero-forcing beamforming}
In ZF beamforming, the matrix $\bar{\bF}_i$ in (\ref{base1}) is the right inverse of the channel matrix $\mathbf{H}_{i,i}^H$:
\begin{equation} \label{ZF_beamf_matrix}
	\bar{\pmb{F}}_i =\Row[\bar{\bff}_{i,k}]_{k\in\clK} = \pmb{H}_{i,i}(\pmb{H}_{i,i}^H\pmb{H}_{i,i})^{-1},
\end{equation}
which exists only when $\pmb{H}_{i,i}^H\pmb{H}_{i,i}$ is nonsingular, particularly requiring
$N > N_{UE}$. It can be seen that
\[
\beta_{i,i}\bH^H_{i,i}\bF_i=\beta_{i,i}\bH^H_{i,i}\pmb{H}_{i,i}(\pmb{H}_{i,i}^H\pmb{H}_{i,i})^{-1}
\mbox{diag}[\sqrt{p_{i,k}}]_{k\in\clK}=\mbox{diag}[\sqrt{\beta_{i,i,k}}\sqrt{p_{i,k}}]_{k\in\clK}
\]
and thus the inter-user interference $\sigma^{\sf U}_{i,k}(\bp_i)$ in (\ref{ue}) is forced to zero. As such,
$\alpha_{i,k}$ defined by (\ref{base3}) is $\beta_{i,i,k}$, while
$\lambda_{i,k}(\bp)$ defined by (\ref{base4}) is
\begin{equation}\label{minf}
\lambda_{i,k}(\bp)=\sigma_{i,k}^{\sf C}(\bp)
\end{equation}
with $\sigma_{i,k}^{\sf C}(\bp)$ defined from (\ref{ICIs}).

From (\ref{channel}) we also define $\bH_{i,i}^{w}\triangleq [\bh^w_{i,i,k}]_{k\in\clK}$ so
\[
\bH_{i,i}=\mathbf{\Theta}_i^{1/2}\bH^w_{i,i}
\]
and $\bH_{i,i}^H\bH_{i,i}=(\bH^w_{i,i})^H\mathbf{\Theta}_i\bH^w_{i,i}$, which has rank not more than $r_i<N$. This makes
matrix $(\bH^w_{i,i})^H\mathbf{\Theta}_i\bH^w_{i,i}$ quicker ill-conditioned as the number $N_{UE}$  of users increases.
We now follow the regularization technique \cite{PHS05,NE08} to consider the following class of
RZF beamforming
\begin{equation}\label{mass2a}
\bar{\bF}_i\triangleq \mathbf{H}_{i,i}(\mathbf{H}_{i,i}^H \mathbf{H}_{i,i} + \eta \mathbf{I}_{2M})^{-1},
\end{equation}
with $\eta>0$. The optimal $\eta$ is not known and we just follow  \cite{PHS05,NE08,Wang2012WCommLett}
to choose
\begin{equation}\label{mass5}
\eta = {2M}\sigma^2/P_i^{\max}.
\end{equation}

Then
\begin{eqnarray}
\beta_{i,i}\mathbf{H}_{i,i}^H\bar{\bF}_i\mbox{diag}[\sqrt{p_{i,k}}]_{k\in\clK}&=&\beta_{i,i}\mathbf{H}_{i,i}^H\mathbf{H}_{i,i}(\mathbf{H}_{i,i}^H \mathbf{H}_{i,i} + \eta \mathbf{I}_{2M})^{-1} \mbox{diag}[\sqrt{p_{i,k}}]_{k\in\clK}\nonumber\\
&=&\beta_{i,i}\mbox{diag}[\sqrt{p_{i,k}}]_{k\in\clK}-\eta \beta_{i,i}(\mathbf{H}_{i,i}^H \mathbf{H}_{i,i} + \eta \mathbf{I}_{2M})^{-1}\mbox{diag}[\sqrt{p_{i,k}}]_{k\in\clK}\nonumber\\
&=&\beta_{i,i}\mbox{diag}[\sqrt{p_{i,k}}]_{k\in\clK}-\eta \beta_{i,i}\mathbf{G}_i(\eta)\mbox{diag}[\sqrt{p_{i,k}}]_{k\in\clK}
\label{mass3a}
\end{eqnarray}
for Hermitian symmetric positive definite matrix
\begin{align} \label{G_pos}
\mathbf{G}_{i}(\eta) = \left[\begin{matrix}\mathbf{g}_{i,1}\cr
...\cr
\mathbf{g}_{i,2M}\end{matrix}\right]
=\left[\begin{matrix}g_{i,1,1}&...&g_{i,1,2M}\cr
...&...&...\cr
g_{i,2M,1}&...&g_{i,2M,2M}\end{matrix}\right]
= (\mathbf{H}_{i,i}^H \mathbf{H}_{i,i} + \eta \mathbf{I}_{2M})^{-1}.
\end{align}
The inter-user interference $\sigma^{\sf U}_{i,k}(\bp_i)$ defined by (\ref{IUIs}) is
\begin{equation}\label{mcin2}
\sigma^{\sf U}_{i,k}(\bp_i)=\eta^2 \beta_{i,k} \sum_{\ell\in\clK\setminus\{k\}}|g_{i,k,\ell}|^2 p_{i,\ell},
\end{equation}
and the transmit power function defined by (\ref{base5}) is defined accordingly.
\subsection{Cell-wide zero-forcing beamforming (CWZF)}
The design of cell-wide ZF (CWZF) beamforming is to ignore the multi-cell interference (\ref{ICIs}), i.e. it aims at optimizing
\begin{equation} \label{rate_cbzf_mul}
r_{i,k}(p_{i,k}) = \ln \left( 1 +  \beta_{i,i,k} p_{i,k}/\sigma^2\right).
\end{equation}
For simplicity of presentation, in this subsection only we use the notation
\begin{equation}\label{barbe1}
\beta_{i,i,k}\rightarrow \bar{\beta}_{i,k}.
\end{equation}
Accordingly, CWZF targets the following  individual EE maximization problems for  cells
$i\in\clI$, ignoring
the intercell-interference (\ref{ICIs}):
\begin{subequations}  \label{CBZF_mcell_opt}
\begin{align}
	\ds\max_{\mathbf{p}_i} & \quad  \ds\frac{\sum_{k\in\clK} \ln \left( 1 + \bar{\beta}_{i,k} p_{i,k} / \sigma^2 \right)} {\pi_i(\mathbf{p}_i)} \label{CBZF_mcell_opta} \\
	\st & \quad \sum_{k\in\clK}||\bar{\bff}_{i,k}||^2 p_{i,k} \leq P_i^{\text{max}} \, , \label{CBZF_mcell_optb} \\
	&  \quad \ln \left( 1 + \bar{\beta}_{i,k}  p_{i,k} / \sigma^2 \right) \geq \hat{r}_{i,k}, \, k\in\clK, \label{CBZF_mcell_optc} 	
\end{align}
\end{subequations}
where $\hat{r}_{i,k}$ is set to be $\hat{r}_{i,k} > \bar{r}_{i,k}$ to compensate the performance loss in the real performance caused by ignoring the intercell-interference (\ref{ICIs}).

Our conference paper \cite{NTDP17} proposed the following treatment  for (\ref{CBZF_mcell_opt}).
First, it follows from (\ref{CBZF_mcell_optc}) that
\[
p_{i,k}\geq \bar{p}_{i,k}:= \sigma^2(e^{\hat{r}_{i,k}}-1)/\bar{\beta}_{i,k},
\]
By making variable change
\[
p_{i,k}=\tilde{p}_{i,k}+\bar{p}_{i,k}
\]
it is straightforward to solve (\ref{CBZF_mcell_opt}) by Dinkelbach's type algorithm, which seeks $t>0$ such that
the optimal solution of the following optimization problem is zero:
\begin{subequations}\label{ee2}
	\begin{align}
	\max_{\tilde{\mathbf{p}}_i}\ & \sum_{k\in\clK} \ln \left( a_{i,k} + \bar{\beta}_{i,k}  \tilde{p}_{i,k} / \sigma^2 \right)
	-t \cdotp \tilde{\pi}_i(\tilde{\mathbf{p}}_i)\label{ee2a} \\
	\st &  \sum_{k\in\clK}||\bar{\bff}_{i,k}||^2\tilde{p}_{i,k} \leq \bar{P}_i^{\text{max}} \, , \, \tilde{p}_{i,k} \geq 0 \, , \, k\in\clK,\label{ee2b}
	\end{align}
\end{subequations}
where
$a_{i,k}=1+\bar{\beta}_{i,k}\bar{p}_{i,k}/\sigma^2$,
$\bar{P}_{i, \rm cir} = \alpha \sum_{k\in\clK} ||\bar{\bff}_{i,k}||^2\bar{p}_{i,k} + P_{\rm cir}$,
$P_{\rm cir} = N P_{a} + P_{c}$,
$\bar{P}_i^{\rm max} = P_i^{\rm max} - \sum_{k\in\clK}||\bar{\bff}_{i,k}||^2 \bar{p}_{i,k}$,
$\tilde{\pi_i}(\tilde{\mathbf{p}}_i) \triangleq \alpha \sum_{k\in\clK} ||\bar{\bff}_{i,k}||^2\tilde{p}_{i,k} + \bar{P}_{i, \rm cir}$.\\
For $t>0$ fixed, problem (\ref{ee2}) admits the optimal solution in closed-form:
\begin{equation} \label{opt_wf}
\tilde{p}_{i,k}^* =\ds \left[ \frac{1}{||\bar{\bff}_{i,k}||^2(t \alpha + \lambda )} - \frac{a_{i,k} \sigma^2}{\bar{\beta}_{i,k}} \right]^+,
k\in\clK.
\end{equation}
Here and after, $[x]^+ = \max\{0,x\}$ and $\lambda  = 0$ whenever
\[
\sum_{k\in\clK} \left[ \frac{1}{||\bar{\bff}_{i,k}||^2 t \alpha} - \frac{a_{i,k} \sigma^2}{\bar{\beta}_{i,k}} \right]^+ \leq \bar{P}_i^{\text{max}}.
\]
Otherwise,  $\lambda >0$ is such that
\begin{align} \label{opt_wf_1}
\sum_{k\in\clK} \left[ \frac{1}{||\bar{\bff}_{i,k}||^2 (t \alpha + \lambda )} - \frac{a_{i,k} \sigma^2}{\bar{\beta}_{i,k}} \right]^+ = \bar{P}_i^{\text{max}},
\end{align}
which can be easily located by the bisection search.

However, in contrast to \cite{NTDP17}, which uses bisection in locating the optimal $t$,
we now propose a path-following Dinkelbach's computational procedure  for (\ref{CBZF_mcell_opt}) as follow:
\begin{itemize}
	\item {\it Initialization.} Solve (\ref{ee2}) for $t=0$. Let
$\tilde{\bp}_i^{(opt)}$ be its optimal solution. Set
\[
\bar{t}=\sum_{k\in\clK}\ln \left( a_{i,k} + \bar{\beta}_{i,k}  \tilde{p}^{(opt)}_{i,k} / \sigma^2 \right)/
	\tilde{\pi}_i(\tilde{\mathbf{p}}^{(opt)}_i).
\]
\item Solve (\ref{ee2}) for $t=\bar{t}$ until its optimal value is zero.
Let $\tilde{\bp}_i^{(opt)}$ be its optimal solution.
Reset $\bar{t}=\sum_{k\in\clK}\ln \left( a_{i,k} + \bar{\beta}_{i,k}  \tilde{p}^{(opt)}_{i,k} / \sigma^2 \right)/
	\tilde{\pi}_i(\tilde{\mathbf{p}}^{(opt)}_i)$.
\end{itemize}
\section{TF-wise zero-forcing and regularized zero-forcing beamforming} \label{FT_Beamf}
It can be seen from (\ref{ICIs}) that compared to the near UEs,
the cell edge UEs suffer not only from worse channel conditions  but also
from the inter-cell interference, which cannot be forced to zero or mitigated.
To tackle this issue of the intercell interference, we propose a scheme involving
two separated transmissions within a time slot. During time-fraction
$0 \leq \tau_1 \leq 1$, BS $1$ transmits signal to serve its near UEs while BS $2$ and BS $3$ transmit signals
to serve their far UEs. During the remaining time-fraction $\tau_2=1-\tau_1$, BS $1$ transmits signal to serve its far UEs
 while BS $2$ and BS $3$ transmit signals   to serve their near UEs. Under this time-fraction (TF)-wise scheme,
the cell-edge UEs are  almost  free from the inter-cell interference
because they are served by their BS when the neighbouring BSs serve their near UEs
and thus need a very small transmission power that causes no interference to  other cells. More importantly,
this TF-wise scheme  allows the individual  BS to serve much larger numbers of UEs within the  time slot.
	
Denote by $\clK_{i,1}$ and $\clK_{i,2}$ the set of those UEs in cell $i$, which are
 served during time-fraction $\tau_1$ and $\tau_2$, respectively. Under the proposed scheme,
\[
\begin{array}{c}
\clK_{1,1}=\clK_{ne}, \clK_{1,2}=\clK_{fa},\\
\clK_{i,1}= \clK_{fa}, \clK_{i,2}=\clK_{ne}, i=2, 3.
\end{array}
\]
The following definitions are used:
\begin{equation}\label{ndef}
\begin{array}{c}
\tau\triangleq (\tau_1, \tau_2),
\bs_i^{[q]}\triangleq\Col[s_{i,k}]_{k\in\clK_{i,q}}, \bfy_i^{[q]}\triangleq\Col[y_{i,k}]_{k\in\clK_{i,q}},\\
\bp_i^{[q]}\triangleq \Col[p_{i,k}]_{k\in\clK_{i,q}}, \bp^{[q]}=[\bp_i^s]_{i\in\clI}, \bn_i^{[q]}=\Row[n_{i,k}]_{k\in\clK_{i,q}},
q=1, 2; i\in\clI,\\
(\bH^{[q]}_{j,i})^H\triangleq \Col[\bh^H_{j,i,k}]_{k\in\clK_{i,q}}.
\end{array}
\end{equation}
As mentioned before, the inter-cell interference is weak in this TF-wise beamforming  and thus can be ignored.
The MIMO equation of signal reception in time-fraction $\tau_q$ is thus
\begin{eqnarray}
\bfy_i^{[q]}
=\beta_{i,i}(\bH^{[q]}_{i,i})^H\bF_i^{[q]}\bs_i^{[q]}+\bn_i^{[q]}.
\label{fmimo2}
\end{eqnarray}
We seek $\bF_i^{[q]}$ is the class of
\begin{equation}\label{mbase1}
\bF_i^{[q]}=\bar{\bF}_i^{[q]}\mbox{diag}[1/\sqrt{p_{i,k}}]_{k\in\clK_{i,q}}
\end{equation}
with predetermined $\bar{\bF}_i^{[q]} \in\mathbf{C}^{N\times M}=\Row[\bar{\bff}_{i,k}]_{k\in\clK_{i,q}}$.

The inter-user interference in time-fraction $\tau_q$ defined as
\begin{equation}\label{fIUIs}
\sigma_{i,k}^{[q]}(\bp_i^{[q]})=\beta_{i,i,k} \sum_{\ell \in \mathcal{K}_{i,q} \setminus \{k\}}|\mathbf{h}_{i,i,k}^H\bar{\bff}_{i,\ell}|^2/p_{i,\ell}, \ell \in \clK_{i,q},
\end{equation}
which is a convex function in $\bp_i^{[q]}$.

The information throughput at  UE $(i,k)$, $k\in\clK_{i,q}$ is $\tau_qr^{[q]}_{i,k}(\bp_i^{[q]})$ with
\begin{equation} \label{finf}
     r^{[q]}_{i,k}(\bp_i^{[q]}) \triangleq \ln \left( 1 + \frac{\beta_{i,i,k} |\Hc{i,i,k}{}^H \bar{\bff}_{i,k}|^2/p_{i,k}}{\sigma_{i,k}^{[q]}(\bp_i^{[q]})  + \sigma^2} \right)=
     \ln \left( 1 + \frac{\alpha_{i,k}/p_{i,k}}{\sigma_{i,k}^{[q]}(\bp_i^{[q]})  + \sigma^2} \right)
\end{equation}
for
\begin{equation}\label{alpha}
\alpha_{i,k}\triangleq \beta_{i,i,k} |\Hc{i,i,k}{}^H \bar{\bff}_{i,k}|^2.
\end{equation}
The transmit beamforming power during  time-fraction $\tau_q$ of each cell is $\tau_q\chi_{i}^{[q]}(\bp_i^{[q]})$ with
\begin{equation}\label{fpower1}
\chi_i^{[q]}(\bp_i^{[q]})\triangleq \sum_{k\in\clK_{i,q}}||\bar{\bff}_{i,k}||^2/p_{i,k},
\end{equation}
which must satisfy the power constraint
\begin{equation}\label{fpower2}
\sum_{q=1}^2\tau_q \chi_i^{[q]}(\bp_i^{[q]}) \leq P_i^{\max}, i\in\clI.
\end{equation}
We also impose additionally the following physical constraints
\begin{equation}\label{fpower3}
||\bar{\bff}_{i,k}||^2/3P_i^{\max}\leq p_{i,k}, (i,k)\in\clI\times\clK
\end{equation}
to substance the fact that it is not possible to transmit an arbitrary high power during time-fractions.\\
The entire power consumption for the downlink transmission  is expressed by
\begin{equation}\label{fpbs}
\pi(\tau, \bp) = \sum_{i\in\clI}  (\alpha \sum_{q=1}^2\tau_q\chi_i^{[q]}(\bp_i^{[q]}) + P_{\rm cir}).
\end{equation}
The EE maximization problem  under QoS constraints and power budget is now formulated as
\begin{subequations} \label{fopt}
\begin{eqnarray}
\ds\max_{\tau, \bp} & \ds\frac{\sum_{q=1}^2\tau_q\sum_{i\in\clI}\sum_{k\in\clK_{i,q}}r^{[q]}_{i,k}(\bp^{[q]}_i)} {\pi(\tau, \bp)}\quad\st\quad  (\ref{fpower2}), (\ref{fpower3}), \label{fopta} \\
& \tau_q r^{[q]}_{i,k}(\bp_i^{[q]}) \geq \overline{r}_{i,k}, \,  i\in\clI, k\in\clK_{i,q}, q=1,2, \label{foptb}\\
& \tau_1\geq 0, \tau_2\geq 0, \tau_1+\tau_2\leq 1.\label{foptc}
\end{eqnarray}
\end{subequations}
To address (\ref{fopt}), introduce the new variable
\begin{equation}\label{change_var_rzf_3cell}
\theta=(\theta_1, \theta_2),
\end{equation}
which satisfies the convex constraints
\begin{equation}\label{cond_var_rzf_3cell}
\tau\theta_1 \geq 1,  (1-\tau)\theta_2\geq 1, \theta_1>0, \theta_2>0.
\end{equation}
The power constraint (\ref{fpower2}) is now
\begin{equation}\label{ad1_rzf_3cell}
\Pi_i(\theta_2, \mathbf{p}_i) \triangleq (1-1/\theta_2)\chi^{[1]}_i(\bp_i^{[1]})
+ \chi^{[2]}_i(\bp_i^{[2]})/\theta_2
\leq P_i^{\max}.
\end{equation}
Problem (\ref{fopt}) is now expressed by
\begin{subequations} \label{d6_rzf_3cell}
	\begin{eqnarray}
\ds\max_{\tau,\theta, \mathbf{p}}\ \Phi(\theta,\bp)/\Pi(\theta_2, \mathbf{p})  \quad
\st \quad (\ref{fpower3}), (\ref{cond_var_rzf_3cell}), (\ref{ad1_rzf_3cell}), \label{d6a_rzf_3cell}\\
\ds r^{[q]}_{i,k}(\bp_i^{[q]})/\theta_q\geq \bar{r}_{i,k}, q=1, 2; i\in\clI; k\in\clK_{i,q},	\label{d6b_rzf_3cell}
	\end{eqnarray}
\end{subequations}
where
\[
\Phi(\theta,\bp)\triangleq \ds\sum_{q=1}^2\frac{1}{\theta_q}\sum_{i\in\clI}\sum_{k \in \mathcal{K}_{i,q}}
r^{[q]}_{i,k}(\bp_i^{[q]})
\]
and
\[
\Pi(\theta_2, \mathbf{p}) = \sum_{i\in\clI}\left( \alpha \cdot \Pi_i(\theta_2, \mathbf{p}_i) + P_{\rm cir} \right).
\]
Let $(\tau^{(n)}, \theta^{(n)}, \mathbf{p}^{(n)})$ be a feasible point for (\ref{d6_rzf_3cell})
found from the $(n-1)$th iteration and
\[
t^{(n)}=\Phi(\theta^{(n)},\bp^{(n)})/\Pi(\theta_2^{(n)}, \mathbf{p}^{(n)}).
\]
 By using inequality (\ref{ineq5}) in the Appendix,
\begin{equation}\label{fopt1}
\Pi_i(\theta_2, \mathbf{p}_i)\leq \Pi_i^{(n)}(\theta_2, \mathbf{p}_i)
\end{equation}
for the convex function
\begin{eqnarray}\label{fopt2}
\Pi_i^{(n)}(\theta_2, \mathbf{p}_i) \triangleq \chi^{[1]}_i(\bp_i^{[1]})
+ \chi^{[2]}_i(\bp_i^{[2]})/\theta_2 +\sum_{k \in \clK_{i,1}} ||\bar{\bff}_{i,k}||^2
\left(p_{i,k}/p^{(n)}_{i,k}+\theta_2/\theta_2^{(n)}-3\right)/p^{(n)}_{i,k}\theta_2^{(n)}.
\end{eqnarray}
Therefore, the nonconvex constraint (\ref{ad1_rzf_3cell}) is innerly approximated by the convex constraint
\begin{equation}\label{fopt3}
\Pi_i^{(n)}(\theta_2, \mathbf{p}_i)\leq P^{\max}_i, i\in\clI.
\end{equation}
To innerly approximate the nonconvex constraint (\ref{d6b_rzf_3cell}) in (\ref{d6_rzf_3cell}),
we apply inequality (\ref{ineq4}) in the Appendix for
\[
x =  p_{i,k}/\alpha_{i,k}, \
y = \sigma_{i,k}^{[q]}(\bp_i^{[q]})  + \sigma^2,
\]
and
\[
\bar{x}= p_{i,k}^{(n)}/\alpha_{i,k},
\bar{y}=\sigma_{i,k}^{[q]}(\bp_i^{q,(n)})  + \sigma^2,
\]
to obtain
\begin{equation}\label{fopt4}
r^{[q]}_{i,k}(\bp_i^{[q]})\geq
r_{i,k}^{q,(n)}(\mathbf{p}_i^{[q]})
\end{equation}
for
\begin{equation} \label{fopt5}
r_{i,k}^{q,(n)}(\mathbf{p}_i^{[q]})=\bar{a}_{i,k}^{(n)} -
\bar{b}_{i,k}^{(n)} p_{i,k}/\alpha_{i,k}  - \bar{c}_{i,k}^{(n)} (\sigma_{i,k}^{[q]}(\bp_i^{[q]})+\sigma^2),
\end{equation}
where
\begin{equation}\label{abcn_rzf}
\begin{array}{c}
0<\bar{a}_{i,k}^{(n)} \triangleq \ds r^{[q]}_{i,k}(\bp_i^{q,(n)}) + 2\alpha_{i,k}/\left(p_{i,k}^{(n)}(\sigma_{i,k}^{[q]}(\bp_i^{q,(n)})  + \sigma^2)+\alpha_{i,k}\right),\\
0<\bar{b}_{i,k}^{(n)} \triangleq \ds (\alpha_{i,k})^2/\left(p_{i,k}^{(n)}(\sigma_{i,k}^{[q]}(\bp_i^{q,(n)})  + \sigma^2)+\alpha_{i,k}\right)p_{i,k}^{(n)},\\
0<\bar{c}_{i,k}^{(n)} \triangleq \ds \alpha_{i,k}/\left(p_{i,k}^{(n)}(\sigma_{i,k}^{[q]}(\bp_i^{q,(n)})  + \sigma^2)+\alpha_{i,k}\right)(\sigma_{i,k}^{[q]}(\bp_i^{q,(n)})  + \sigma^2).
\end{array}
\end{equation}
The nonconvex constraint (\ref{d6b_rzf_3cell}) is thus innerly approximated by the following convex constraint:
\begin{equation}\label{cond_rate_1_rzf_3cell}
r_{i,k}^{q,(n)}(\mathbf{p}_i^{[q]}) \geq \theta_q \overline{r}_{i,k}  \, , \, q=1, 2; i\in\clI, k\in\clK_{i,q}.
\end{equation}
Next, we address the terms in the numerator of the objective in (\ref{d6a_rzf_3cell}). By using inequality (\ref{ineq3}) in the Appendix for
\[
x = p_{i,k}/\alpha_{i,k},\
y =  \sigma_{i,k}^{[q]}(\bp_i^{[q]})+\sigma^2,
t \triangleq \theta_q,
\]
and
\[
\bar{x}= p_{i,k}^{(n)}/\alpha_{i,k},
\bar{y}= \sigma_{i,k}^{[q]}(\bp_i^{q,(n)})+\sigma^2,
\bar{t} =\theta_q^{(n)},
\]
we obtain
\begin{equation}\label{fopt6}
r^{[q]}_{i,k}(\bp_i^{[q]})/\theta_q\geq g_{i,k}^{q,(n)}(\theta_q,\mathbf{p}),
\end{equation}
where
\begin{equation} \label{obj1a_rzf_3cell}
g_{i,k}^{q,(n)}(\theta_q,\mathbf{p})\triangleq
a_{i,k}^{(n)} - b_{i,k}^{(n)} p_{i,k}/\alpha_{i,k}  - c_{i,k}^{(n)}(\sigma_{i,k}^{[q]}(\bp_i^{[q]})+\sigma^2) - d_{i,k}^{(n)} \theta_q
\end{equation}
with
\begin{equation}\label{obj1_rzf_1b}
\begin{array}{lll}
0<a_{i,k}^{(n)} &\triangleq& 2r_{i,k}(\bp_i^{q,(n)})/\theta_q^{(n)}
+ 2\alpha_{i,k}/\left(p_{i,k}^{(n)}(\sigma_{i,k}^{[q]}(\bp_i^{q,(n)})+\sigma^2)+\alpha_{i,k}\right)\theta_q^{(n)},\\
0<b_{i,k}^{(n)} &\triangleq& \ds (\alpha_{i,k})^2/\left(p_{i,k}^{(n)}(\sigma_{i,k}^{[q]}(\bp_i^{q,(n)})+\sigma^2)+\alpha_{i,k}\right)p_{i,k}^{(n)} \theta_q^{(n)},\\
0<c_{i,k}^{(n)} &\triangleq& \ds \alpha_{i,k}/\left(p_{i,k}^{(n)}(\sigma_{i,k}^{[q]}(\bp_i^{q,(n)})+\sigma^2)+\alpha_{i,k}\right)
(\sigma_{i,k}^{[q]}(\bp_i^{q,(n)})+\sigma^2) \theta_q^{(n)}, \\
0<d_{i,k}^{(n)} &\triangleq& r_{i,k}(\bp_i^{q,(n)})/(\theta_q^{(n)})^2.
\end{array}
\end{equation}
At the $n$th iteration, the following convex program is solved to generate the next feasible point $(\tau^{(n+1)}, \theta^{(n+1)}, \mathbf{p}^{(n+1)})$ for (\ref{d6_rzf_3cell}):
\begin{eqnarray} \label{d8_rzf_3cell}
\ds\max_{\theta,\tau, \mathbf{p}}\ \ds\sum_{q=1}^2\sum_{i\in\clI}\sum_{k\in\clK_{i,q}}
g_{i,k}^{q,(n)}(\theta_q,\mathbf{p}) - t^{(n)}\sum_{i\in\clI}\left(\alpha\cdot\Pi_i^{(n)}(\theta_2,\bp_i)+P_{\rm cir}\right) \nonumber\\
\st  \quad (\ref{fpower3}), (\ref{cond_var_rzf_3cell}), (\ref{fopt3}), (\ref{cond_rate_1_rzf_3cell}).
\end{eqnarray}

In Algorithm \ref{alg2}, we propose a path-following computational procedure for the EE maximization problem (\ref{d6_rzf_3cell}).

To find  an initial point $(\theta^{(0)}, \mathbf{p}^{(0)})$ for (\ref{d6_rzf_3cell}) we fix $\theta^{(0)}$ such that it satisfies (\ref{cond_var_rzf_3cell}), and solve the
following linear programming problem:
\begin{subequations} \label{feas1_rzf_3cell}
\begin{eqnarray}
	\ds \min_{\mathbf{p}} \ \tilde{\pi}(\bp) \quad
	\st \quad \tilde{\pi}_i(\bp_i)  \leq P_i^{\max}, i\in\clI, \label{feas1b_rzf_3cell} \\
	\ds \alpha_{i,k} p_{i,k} \geq (e^{\theta_q^{(0)}\overline{r}_{i,k}}-1) (\tilde{\sigma}_{i,k}^{[q]}(\bp_i^{[q]})+ \sigma^2), q=1, 2; i\in\clI, k\in\clK_{i,q}, \label{feas1c_rzf_3cell}
	\end{eqnarray}
\end{subequations}
where
\[
\begin{array}{c}
\tilde{\pi}_i(\bp_i)\triangleq \ds(1-\frac{1}{\theta_2^{(0)}})\sum_{k\in\clK_{i,1}}||\bar{\bff}_{i,k}||^2p_{i,k}+
\sum_{k\in\clK_{i,2}}||\bar{\bff}_{i,k}||^2p_{i,k}, i\in\clI,\\
\tilde{\pi}(\bp)\triangleq \sum_{i\in\clI}\tilde{\pi}_i(\bp_i),\\
\tilde{\sigma}_{i,k}^{[q]}(\bp_i^{[q]})\triangleq \beta_{i,i,k} \sum_{\ell \in \mathcal{K}_{i,q} \setminus \{k\}}|\mathbf{h}_{i,i,k}^H\bar{\bff}_{i,\ell}|^2p_{i,k}, k\in\clK_{i,q},
\end{array}
\]
which are linear functions. Note that the linear constraint (\ref{feas1c_rzf_3cell}) represents the following
QoS constraints
\begin{equation}\label{lll}
\frac{1}{\theta_q^{(0)}}\ln \left(1+\frac{\alpha_i p_{i,k}}{\tilde{\sigma}_{i,k}^{[q]}(\bp_i^{[q]})+ \sigma^2}\right) \geq \overline{r}_{i,k}, q=1, 2; i\in\clI, k\in\clK_{i,q}.
\end{equation}
Suppose $\bar{\bp}$ is the optimal solution of (\ref{feas1_rzf_3cell}). Then
an initial point $(\theta^{(0)}, \mathbf{p}^{(0)})$ for (\ref{d6_rzf_3cell}) is $p^{(0)}_{i,k}=1/\bar{p}_{i,k}$.

\begin{algorithm}
	\caption{: Path-following algorithm for solving problem (\ref{d6_rzf_3cell})} \label{alg2}
	\begin{algorithmic}[1]
		\State \textbf{Initialization}: Solve (\ref{feas1_rzf_3cell}) to take its optimal solution as
 a feasible point $(\theta^{(0)}, \mathbf{p}^{(0)})$ for (\ref{d6_rzf_3cell}). Set $n:=0$ and $t^{(0)}:=\Phi(\theta^{(0)},\bp^{(0)})/\Pi(\theta_2^{(0)},\bp^{(0)})$.
		\State \textbf{Repeat}
		\State \quad Solve the problem (\ref{d8_rzf_3cell}) for its optimal solution $(\tau^{(n+1)},\theta^{(n+1)}, \mathbf{p}^{(n+1)})$. Set $t^{(n+1)}:=\Phi(\theta^{(n+1)},\bp^{(n+1)})/\Pi(\theta_2^{(n+1)},\bp^{(n+1)})$.
		\State Set $n:=n+1$.
		\State \textbf{Until} convergence of the objective in (\ref{d6_rzf_3cell}).
	\end{algorithmic}
\end{algorithm}
Similar to Algorithm \ref{alg1}, at least Algorithm \ref{alg2} converges to a locally optimal solution of (\ref{d6_rzf_3cell}) satisfying the KKT conditions of optimality.

For TF-wise  ZF beamforming, $\bar{\bF}^{[q]}_i$ in (\ref{mbase1}) is
the right inverse of the matrix $(\mathbf{H}^{[q]}_{i,i})^H$:
\begin{align} \label{zf1}
	\bar{\pmb{F}}^{[q]}_i =\Col[\bar{\bff}_{i,k}]_{k\in\clK_{i,q}} & = \pmb{H}^{[q]}_{i,i}((\pmb{H}^{[q]}_{i,i})^H\pmb{H}^{[q]}_{i,i})^{-1}.
\end{align}
under which the inter-user interference $\sigma_{i,k}^{q}(\bp_i^{[q]})$ in (\ref{fIUIs}) is zero.

On the other hand, for TF-wise  RZF beamforming, $\bar{\bF}^{[q]}_i$ in (\ref{mbase1}) is
\begin{equation}\label{fmass2a}
\bar{\bF}^{q}_i=\mathbf{H}^{[q]}_{i,i}((\mathbf{H}^{[q]}_{i,i})^H \mathbf{H}^{[q]}_{i,i} + \eta \mathbf{I}_{M})^{-1}.
\end{equation}
with
\begin{equation}\label{fmass5}
\eta = {M}\sigma^2/P_i^{\max}.
\end{equation}
Then
\begin{eqnarray}
\beta_{i,i}(\mathbf{H}^{[q]}_{i,i})^H\bF_i^{[q]}&=&\beta_{i,i}\mbox{diag}[\sqrt{p_{i,k}}]_{k\in\clK_{i,q}}-\eta \beta_{i,i}\mathbf{G}^{[q]}_{i}(\eta)\mbox{diag}[1/\sqrt{p_{i,k}}]_{k\in\clK_{i,q}}
\label{fmass3a}
\end{eqnarray}
for the  Hermitian symmetric positive definite matrix
\begin{align} \label{fG_pos}
\mathbf{G}^{[q]}_{i}(\eta) = \left[\begin{matrix}\mathbf{g}^{[q]}_{i,1}\cr
...\cr
\mathbf{g}^{[q]}_{i,M}\end{matrix}\right]
=\left[\begin{matrix}g_{i,1,1}&...&g_{i,1,M}\cr
...&...&...\cr
g_{i,M,1}&...&g_{i,M,M}\end{matrix}\right]
= ((\mathbf{H}^{[q]}_{i,i})^H \mathbf{H}^{[q]}_{i,i} + \eta \mathbf{I}_{M})^{-1}.
\end{align}
In this case, $\alpha_{i,k}$ defined by (\ref{alpha}) is
\[
\alpha_{i,k}=\beta_{i,k}(1-\eta g_{i,k,k})^2,
\]
while the inter-user interference $\sigma_{i,k}^{[q]}(\bp_i^{[q]} )$ in (\ref{fIUIs})
is
\begin{equation}\label{fIUIa}
\sigma_{i,k}^{[q]}(\bp_i^{[q]})\triangleq \eta^2 \beta_{i,k} \sum_{\ell \in \clK_{i,q} \setminus \{k\}}|g_{i,k,\ell}|^2/ p_{i,\ell},
k \in \clK_{i,q}.
\end{equation}
The transmit power function $\chi_i^{[q]}(\bp_i^{[q]})$ defined by
(\ref{fpower1}) is also represented as
\begin{equation}\label{d3_rzf_3cell}
\chi_i^{[q]}(\bp_i^{[q]})=
 \mbox{trace}\left(\mathbf{G}^{[q]}_{i}(\eta)
(\mathbf{H}^{[q]}_{i,i})^H \mathbf{H}^{[q]}_{i,i} \mathbf{G}^{[q]}_{i}(\eta) \mbox{diag}[1/p_{i,k}]_{k\in\clK_{i,q}} \right).
\end{equation}

\section{Numerical Simulations}
In this section, we evaluate the performance of the proposed algorithms by numerical examples for different scenarios of single-cell, two-cell and three-cell networks.
Unless otherwise stated, it is assumed that $N_{\sf ne}=N_{\sf fa}=N_{UE}/2$. The cell-edge UEs
are equally distributed at the cell boundaries, while the near UEs are equally distributed nearly the BSs.
Each of BSs is located at the centre of a hexagon cell with radius $1$ km and
equipped with an $8\times 8$ uniform planar array (UPA) of antennas ($8$ rows in the horizontal dimension and $8$ columns in the vertical dimension). Thus, the total number of antennas at each BS is $N=64$.
 A popular model for the spatial correlation matrix $\Theta_j$ in (\ref{channel}) is an $2D$ extension  \cite{Adhi_TIT_2013, AdhikaryJSAC_2015}
of one ring model \cite{Shetal00}, which is of very low rank \cite{STND17} under the standard assumption  that
antennas are a half-wavelength spaced to result in a form factor of $0.25$ m $\times$ $0.25$ m \cite{Ketal14}. To investigate the impact of the spatial
correlation to the number of UEs as well as the users's QoS that massive MIMO can promise, we adopt
the standard exponential correlation model, where the correlation between antenna $(p,q)$ and antenna $(m,n)$ is modelled by
\begin{align}
[\mathbf{\Theta}]_{(p,q),(m,n)} = \rho^{|p-m|+|q-n|}
\end{align}
with $0 < \rho < 1$, which was also used e.g. in \cite{Wagner2012}.
To study the effect of spatial correlation to capacity of massive MIMO,
we consider two cases of $\rho = 0.9$ and $\rho = 0.5$, which  correspond to  high  and medium
spatial correlations.

Other simulation parameters for generating large scale fading   in Table I are similar to those used in \cite{Bjornson2013}.
The throughput threshold for all users is set as $\bar{r}_{i,k}\equiv r \in\{ 0.4, 1\}$ bps/Hz \cite[Table I]{Andrews2014}.
\begin{table}[ht]
	\centering
	\caption{Large scale fading Setup}
	\begin{tabular}{ |p{4cm}||p{4.5cm}| }
		\hline
		Parameter & Numerical value \\
		\hline
		Carrier frequency / Bandwidth   		& $2$GHz / $10$MHz  \\
		BS transmission power					& $46$ dBm \\
		Path loss from BS to UE				& $128.1 + 37.6\loga_{10}R$ [dB], R in km \\
		Shadowing standard deviation 			& $8$ dB \\
		Noise power density 					& $-174$ dBm/Hz \\
		Noise figure 							& $9$ dB \\		
		Drain efficiency of amplifier 			& $\alpha = 1/0.388$ \\
		Circuit power per antenna 			& $P_A$ = $189$ mW \\
		Non-transmission power 				& $P_{C}$ = $40$ dBm \\
		\hline
	\end{tabular}
\end{table}

\subsection{Single-cell network}
\begin{figure}[H]
	\centering
	\centerline{\includegraphics[width=0.7 \textwidth]{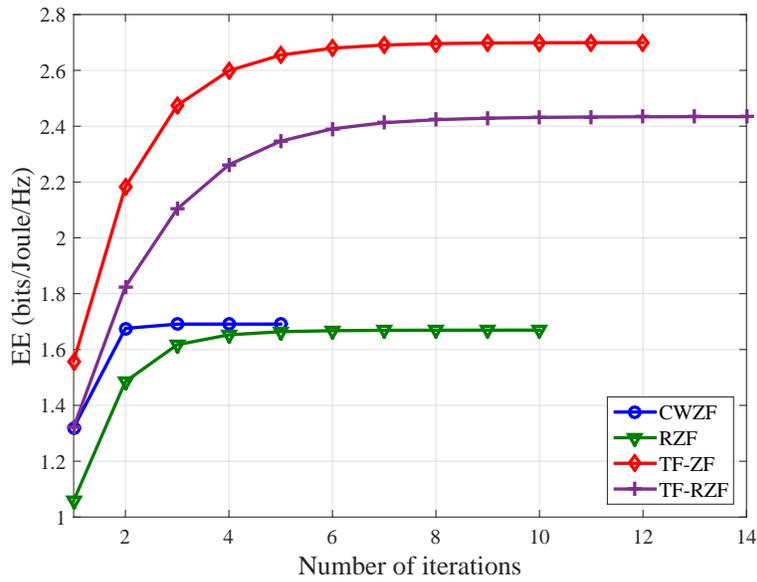}}
	\caption{The convergence of {CWZF}, {RZF}, {TF-ZF} and {TF-RZF} vs. iteration number under  $N_{\sf UE} = 40$, $\rho = 0.9$ and $r = 0.4$ bps/Hz.}
	\label{fig:1_1cell_conv}
\end{figure}

A typical convergence of the proposed
Algorithm \ref{alg1} for RZF beamforming,  Dinkelbach's type iterations for CWZF beamforming
and Algorithm \ref{alg2} for TF-based ZF and RZF beamforming is provided
by Fig. \ref{fig:1_1cell_conv}, where all of them are seen to converge rapidly within several iterations.
It is worthy to mention that the new path-following Dinkelbach's iterations converge much more rapidly than
that proposed in \cite{NTDP17}, which are based on bisection for locating the optimal value of $t$
in (\ref{ee2}).

\begin{figure}[H]
	\centering
	\centerline{\includegraphics[width=0.75 \textwidth]{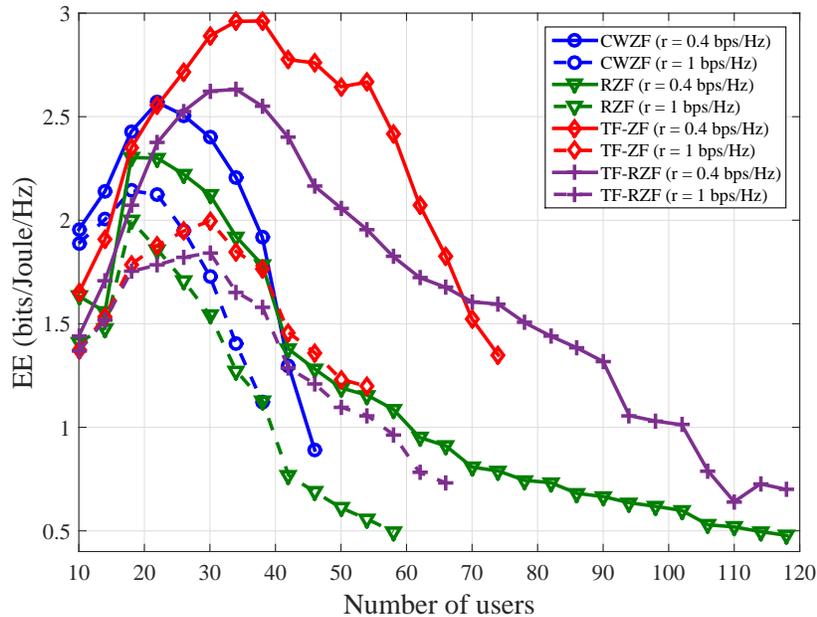}}
	\caption{The EE performance in {CWZF}, {RZF}, {TF-wise  ZF} and {TF-wise  RZF} vs. the number of users under $\rho = 0.9$ and $r \in \{0.4, 1\}$ bps/Hz.}
	\label{fig:2_EE_NoUE_1cell}
\end{figure}
Fig. \ref{fig:2_EE_NoUE_1cell} plots the EE performance of the proposed beamforming approaches versus the number of users under $\rho = 0.9$.
RZF beamforming is always capable of serving a much larger numbers of UEs than ZF beamforming is.
For the throughput threshold $r = 0.4$ bps/Hz ($r = 1$ bps/Hz, resp.), CWZF beamforming and TF-wise  ZF beamforming cannot serve more than $46$ UEs ($38$ UEs, resp.) and $82$ UEs ($54$ UEs, resp.). Meanwhile,
both RZF beamforming and TF-wise  RZF beamforming can serve up to  $120$ UEs
($66$ UEs, resp.) for $r = 0.4$ bps/Hz ($r = 1$ bps/Hz, resp.) but the latter clearly outperforms the
former in term of EE. Note that both numbers $120$ and $66$ of the served UEs excess the number $64$ of BS's antennas. Both
optimal time-fraction allocation for two separated transmission within the time slot and optimal power allocation for
beamformers enable massive MIMO to serve numbers of UEs that  are larger than the number of transmit antennas.

\begin{figure}[H]
	\centering
	\centerline{\includegraphics[width=0.75 \textwidth]{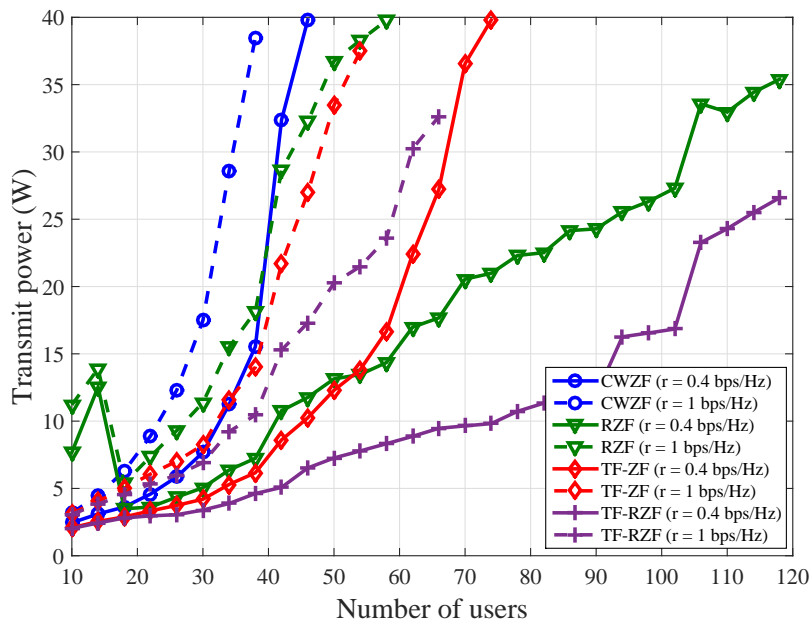}}
	\caption{The transmit power in {CWZF}, {RZF}, {TF-wise  ZF} and {TF-wise  RZF} vs. the number of users under $\rho = 0.9$ and $r \in \{0.4, 1\}$ bps/Hz.}
	\label{fig:2_Pow_NoUE_1cell}
\end{figure}
Furthermore, all EE performances  increase quickly to a certain value of
$N_{UE}$ and drop after that. Fig. \ref{fig:2_Pow_NoUE_1cell} reveals that this drop is caused
by the increased  total transmit power.
There is no magic number $N_{UE}$, under which all the EE performances attain their peak.
Of course, increasing the throughput threshold from $0.4$ bps/Hz to $1$ bps/Hz leads to decreasing
numbers of the served UEs and degrading EE performance.
Fig. \ref{fig:2_Pow_NoUE_1cell} also shows that TF-wise  beamforming could manage the power control better than
other beamforming schemes.

\begin{figure}[H]
	\centering
	\centerline{\includegraphics[width=0.75 \textwidth]{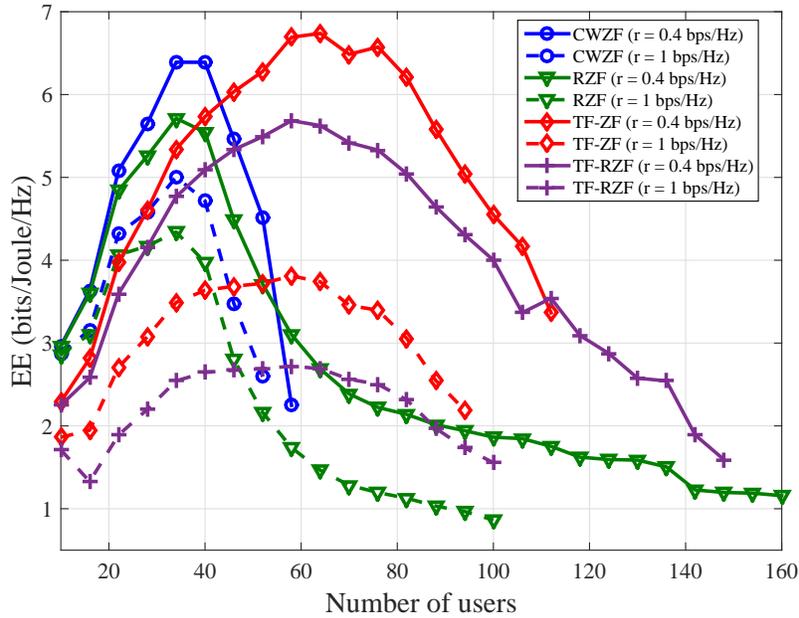}}
	\caption{The EE performance in {CWZF}, {RZF}, {TF-wise  ZF} and {TF-wise  RZF} vs. the number of users under $\rho = 0.5$ and $r = \{0.4, 1\}$ bps/Hz.}
	\label{fig:2_EE_NoUE_1cell_1}
\end{figure}

\begin{figure}[H]
	\centering
	\includegraphics[width=0.75 \textwidth]{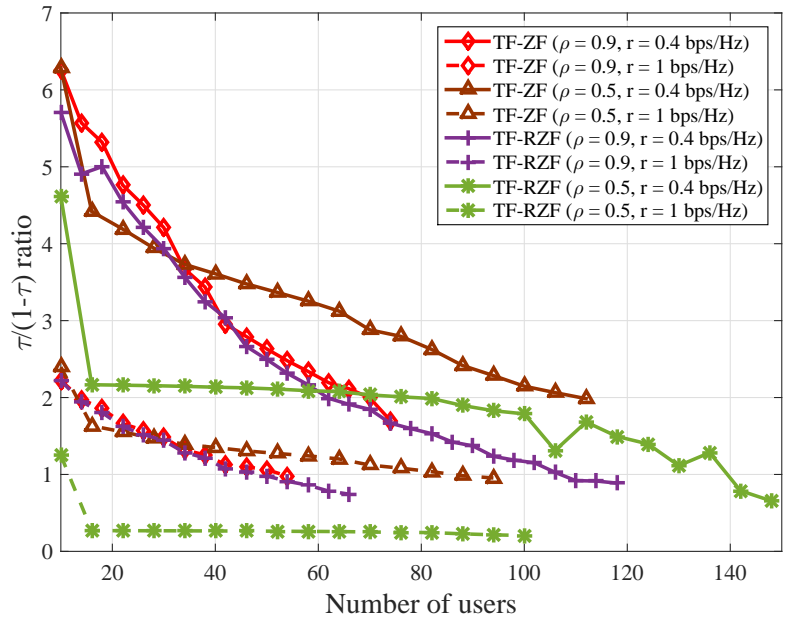}	
	\caption{The value of $\tau/(1-\tau)$ in {TF-ZF} and {TF-RZF} vs. the number of users.}
	\label{fig:ratio_tau_NoUE_1cell}
\end{figure}
Fig. \ref{fig:2_EE_NoUE_1cell_1} plots the EE performance of the proposed beamforming schemes
under  $\rho = 0.5$. Lower spatial correlation obviously leads to not only
better EE but also larger numbers of the served UEs. Specifically, the EE performance is doubly increased
in all proposed beamforming schemes and  TF-wise  RZF beamforming can serve $160$ UEs vs $120$ UEs served
under $\rho=0.9$.

\begin{figure}[H]
	\centering
	\includegraphics[width=0.75 \textwidth]{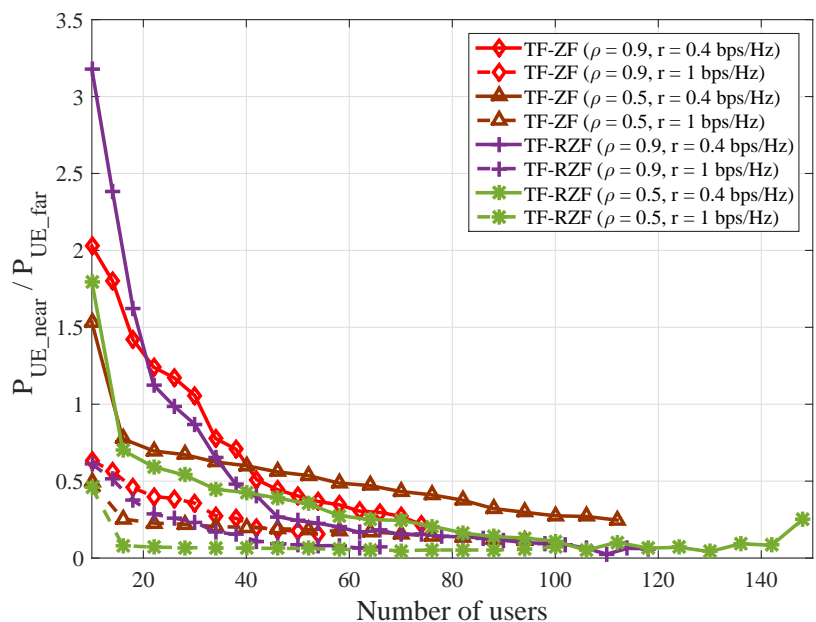}	
	\caption{The ratio of transmit power for near UEs and far UEs in {TF-ZF} and {TF-RZF} vs. the number of users.}
	\label{fig:ratio_pow_NoUE_1cell}
\end{figure}

Fig. \ref{fig:ratio_tau_NoUE_1cell}  and Fig. \ref{fig:ratio_pow_NoUE_1cell} plot the ratio between time-fractions
in serving the near UEs and the cell-edge UEs
and the corresponding power ratio, which are monotonically
decreased in the total number $N_{UE}$ of UEs. Recalling that $N_{\sf ne}=N_{\sf fa}=N_{UE}/2$ in our setting,
at small $N_{UE}$ / small $N_{\sf fa}$ more time-fraction and power are allocated to the near UEs to maximize
their throughput. On the other hand, at large $N_{UE}$ / large $N_{\sf fa}$, more time-fraction and power
must be allocated to the far UEs  in assuring  their QoS.

\subsection{Two-cell network}
\begin{figure}[H]
	\centering
	\centerline{\includegraphics[width=14cm,height=7cm]{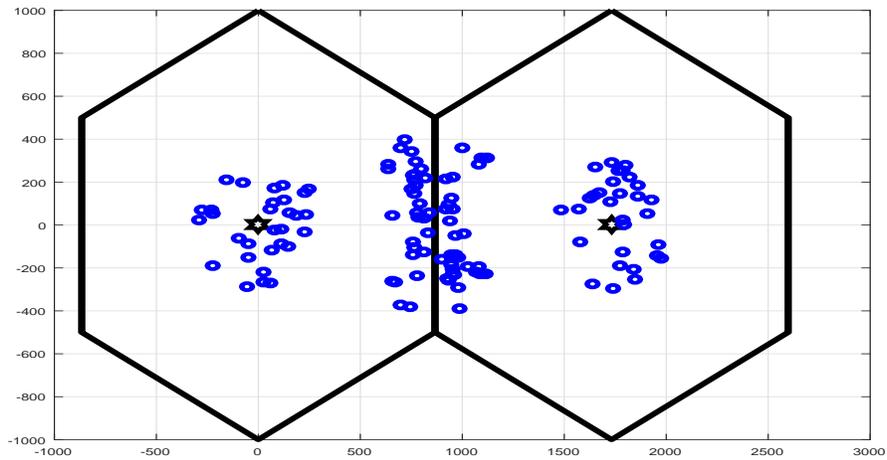}}
	\caption{An equally mixed-coupled two-cell scenario. Each cell has a total of 60 UEs.}
	\label{fig:0_SM_2cell}
\end{figure}

The network is depicted by  Fig. \ref{fig:0_SM_2cell}, where the cell-edge UEs are located
at the boundary areas  between the cells.
Under the TF-wise  beamforming schemes,
during time-fraction $0 \leq \tau \leq 1$, BS $1$ serves its near UEs while BS $2$ serves its cell-edge UEs.
During the remaining fraction ($1-\tau$), BS $1$ serves its cell-edge UEs while BS $2$ serves its near UEs. The cell-edge
UEs are thus free from the inter-cell interference.

\begin{figure}[H]
	\centering
	\includegraphics[width=0.75 \textwidth]{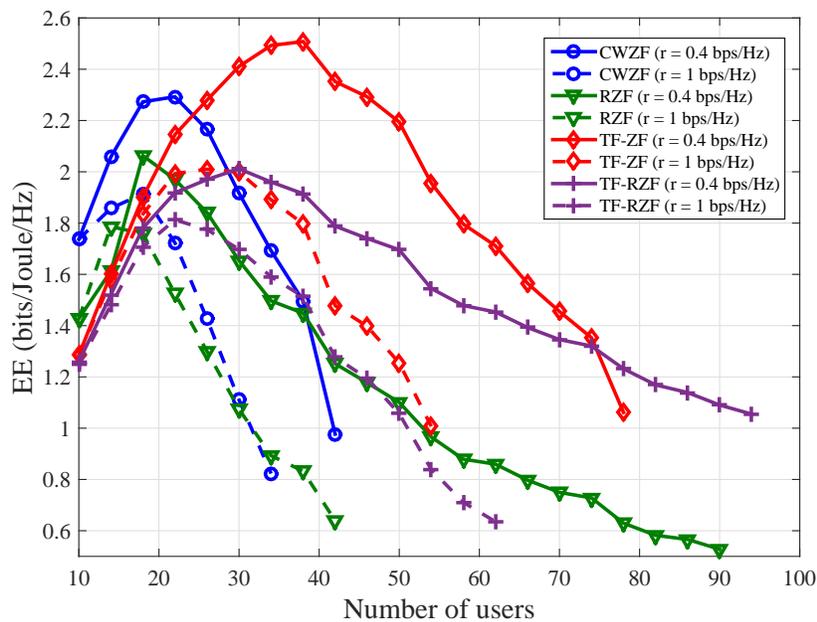}
	\caption{The EE performance in {CWZF}, {RZF} and {TF} vs. the number of users under $\rho = 0.9$, $r \in \{0.4, 1\}$ bps/Hz and $\hat{r} = \{0.6, 1.4\}$ bps/Hz.}
	\label{fig:EE_NoUE_2cell}
\end{figure}
\begin{figure}[H]
	\centering
	\includegraphics[width=0.72 \textwidth]{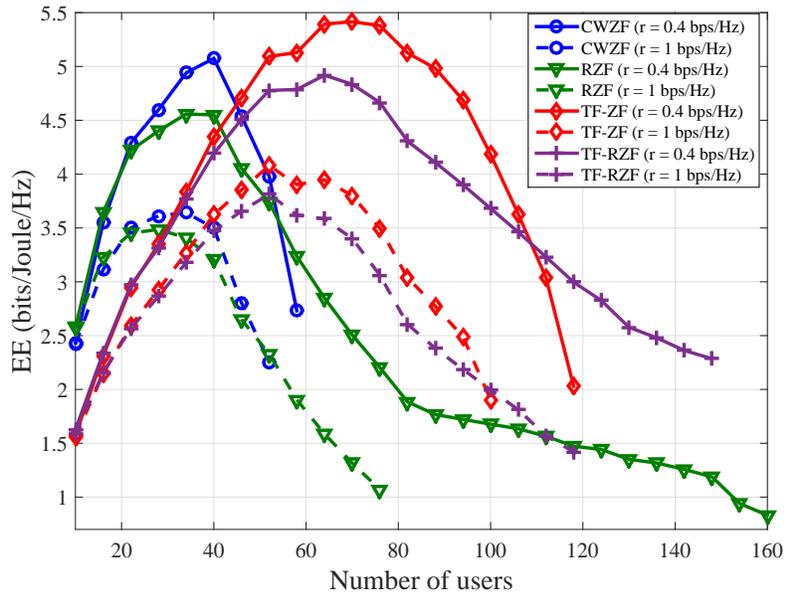}
	\caption{The EE performance in {CWZF}, {RZF} and {TF} vs. the number of users under $\rho = 0.5$, $r \in \{0.4, 1\}$ bps/Hz and $\hat{r} = \{0.6, 1.4\}$ bps/Hz.}
	\label{fig:EE_NoUE_2cell_1}
\end{figure}

Fig. \ref{fig:EE_NoUE_2cell} and Fig. \ref{fig:EE_NoUE_2cell_1} show the superior performance of TF-wise   beamforming
schemes over others.
For the throughput threshold $r = 0.4$ bps/Hz, CWZF beamforming cannot serve more than $40$ UEs and $60$ UEs
while TF-wise   ZF beamforming still serves up to $80$ UEs and $120$ UEs,  respectively.
Under both spatial correlation degrees, RZF beamforming and TF-wise   RZF beamforming
can serve up to $90$ UEs and $150$ UEs but the latter  significantly outperforms the former in term of EE.
It is observed that the EE gap in assuring the throughput thresholds becomes wider as the number $N_{UE}$ of UEs
increases.

\begin{figure}[H]
	\centering
		\includegraphics[width=0.72 \textwidth]{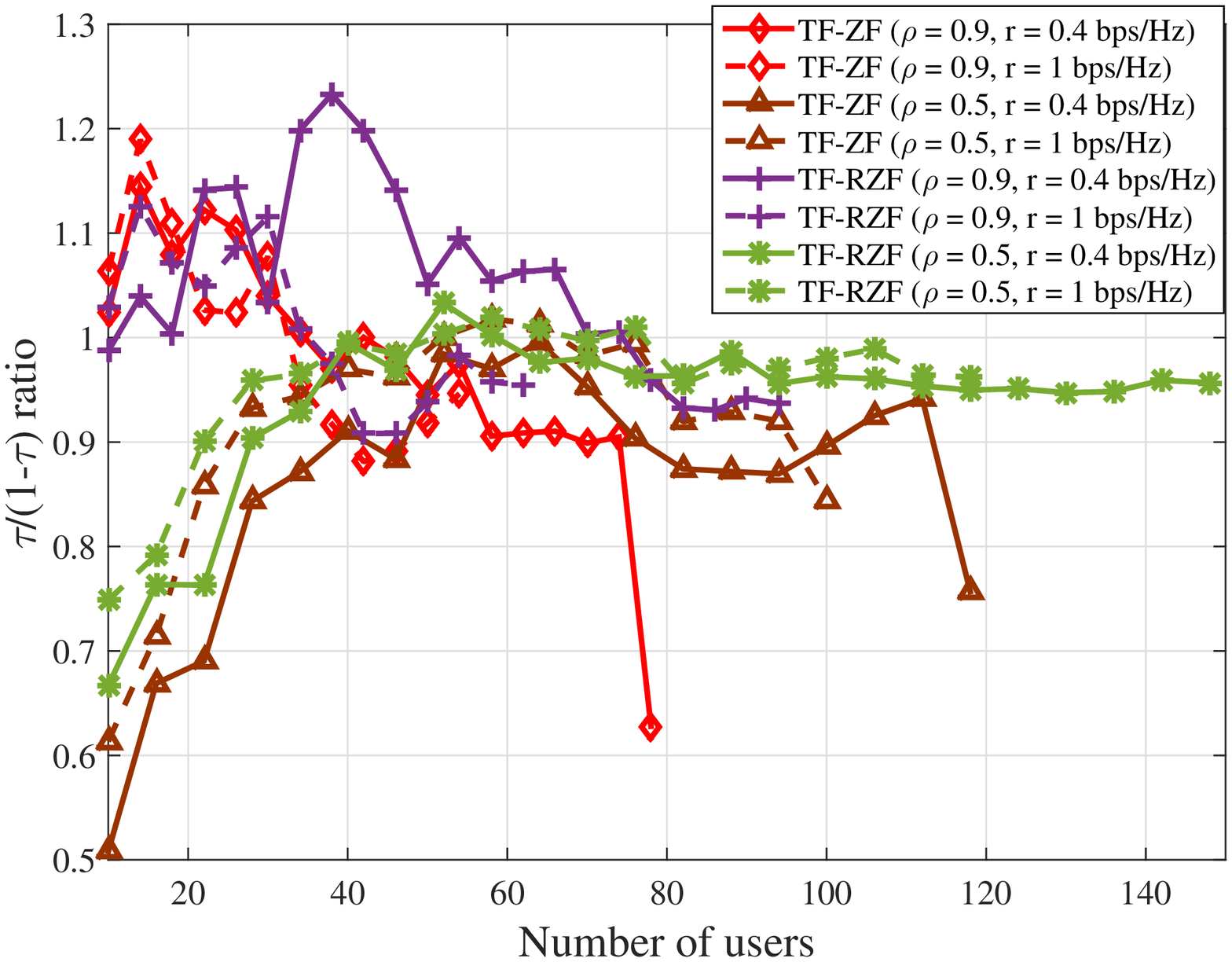}	
	\caption{The value of $\tau/(1-\tau)$ vs. the number of users.}
	\label{fig:ratio_tau_NoUE_2cell}
\end{figure}

\begin{figure}[H]
	\centering
	\includegraphics[width=0.72 \textwidth]{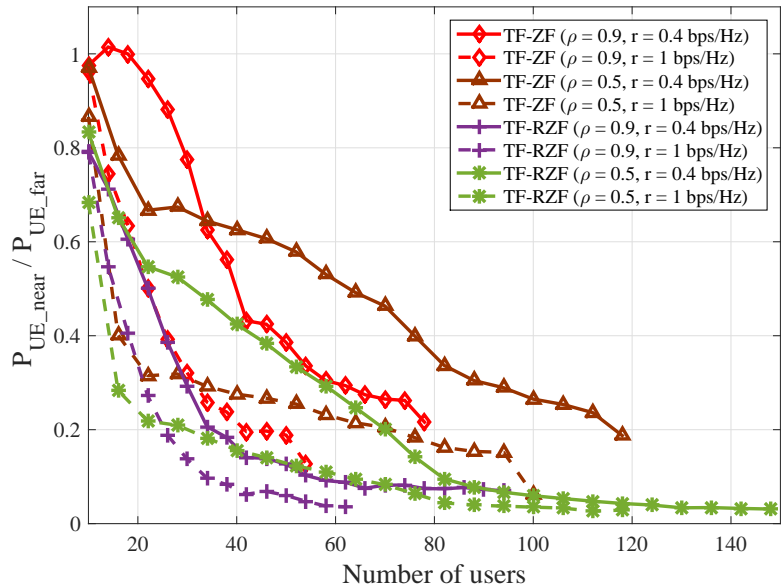}	
	\caption{The total transmit power for near UEs and far UEs ratio vs. the number of users.}
	\label{fig:ratio_pow_NoUE_2cell}
\end{figure}
Interestingly, Fig. \ref{fig:ratio_tau_NoUE_2cell} and Fig. \ref{fig:ratio_pow_NoUE_2cell} show that the time-fraction allocation and power allocation in this two-cell case are quite different from that in the single-cell case. They are more or less balanced because the same numbers of cell-edge  UEs and near  UEs are served in different time-fractions.

\subsection{Three-cell network}
\begin{figure}[H]
	\centering
	\includegraphics[width=0.75 \textwidth]{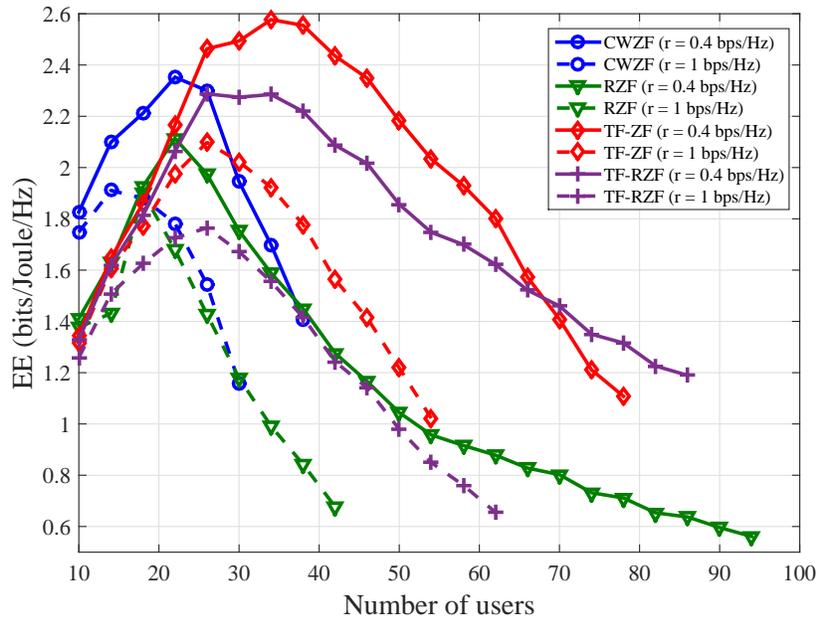}
	\caption{The EE performance in {CWZF}, {RZF} and {TF} vs. the number of users under $\rho = 0.9$, $r = \{0.4, 1\}$ bps/Hz and $\hat{r} = \{0.6, 1.4\}$ bps/Hz.}
	\label{fig:EE_NoUE_3cell}
\end{figure}

\begin{figure}[H]
	\centering
	\includegraphics[width=0.75 \textwidth]{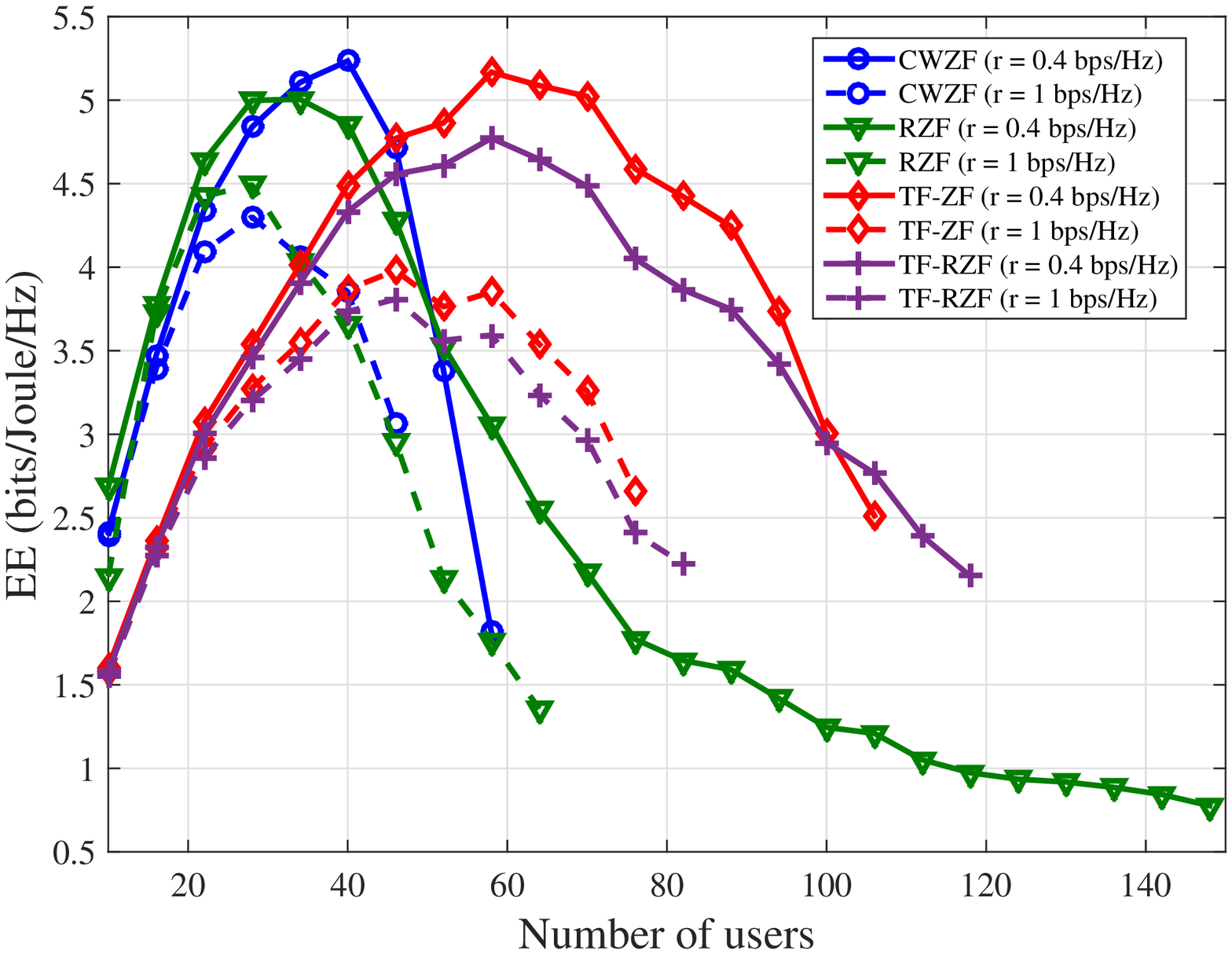}
	\caption{The EE performance in {CWZF}, {RZF} and {TF} vs. the number of users under $\rho = 0.5$, $r = \{0.4, 1\}$ bps/Hz and $\hat{r} = \{0.6, 1.4\}$ bps/Hz.}
	\label{fig:EE_NoUE_3cell_1}
\end{figure}

We return to a three-cell network illustrated by  Fig. \ref{fig:0_SM_3cell}. Being free from inter-cell interference,
TF-wise   beamforming schemes can serve  higher numbers of UEs with higher EE achieved, as
Fig. \ref{fig:EE_NoUE_3cell} and Fig. \ref{fig:EE_NoUE_3cell_1} show. Particularly,
TF-wise   ZF beamforming and TF-wise RZF beamforming are able to serve at least $80$ UEs  and $120$ UEs per cell  for
$\rho=0.9$ and $\rho=0.5$, respectively. Both
RZF beamforming and TF-wise   RZF beamforming can serve up to $150$ UEs for $\rho=0.5$
but the latter clearly outperform the former in terms of EE.

\begin{figure}[H]
	\centering
		\includegraphics[width=0.75 \textwidth]{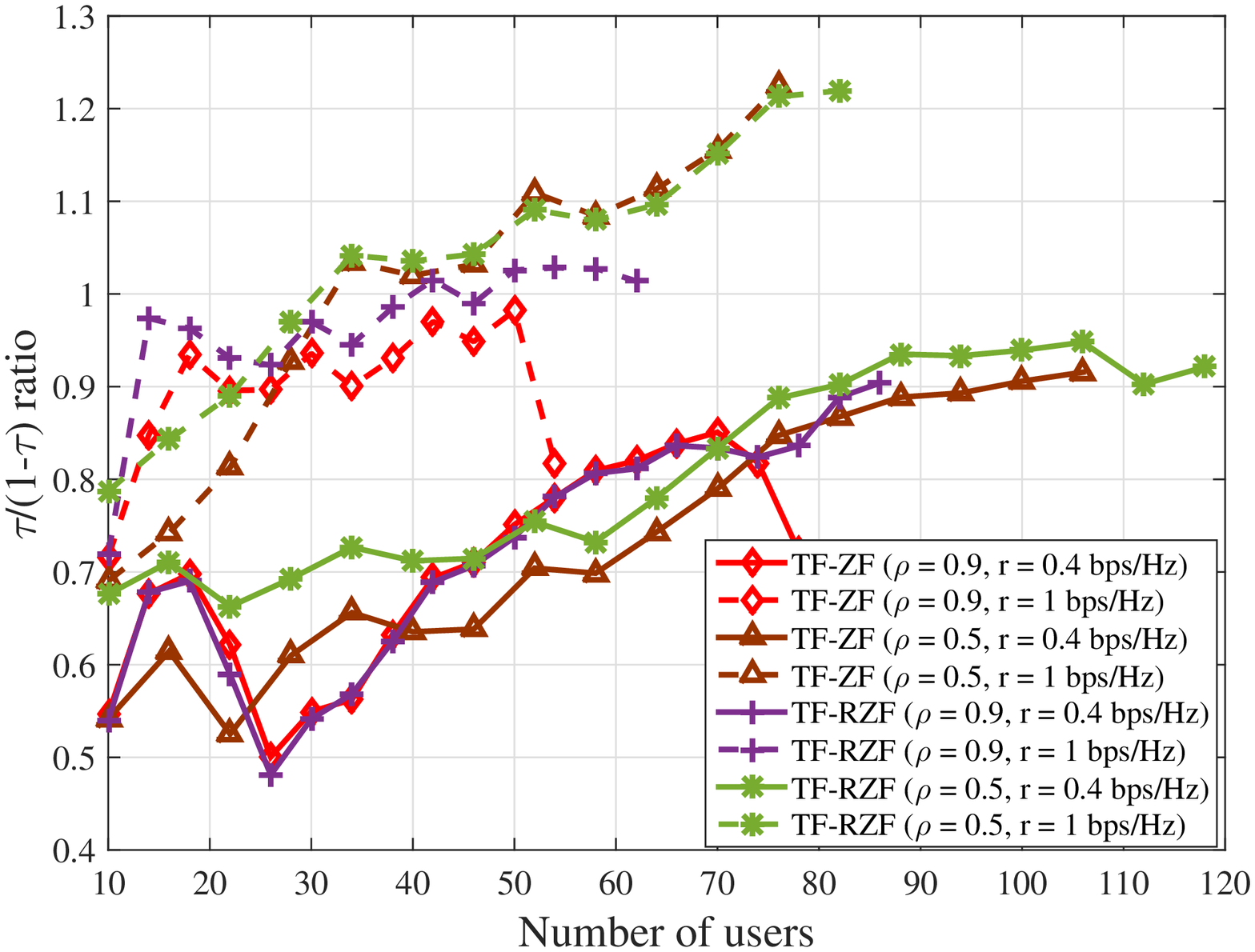}	
	\caption{The value of $\tau/(1-\tau)$ vs. the number of users.}
	\label{fig:ratio_tau_NoUE_3cell}
\end{figure}

\begin{figure}[H]
	\centering
	\includegraphics[width=0.75 \textwidth]{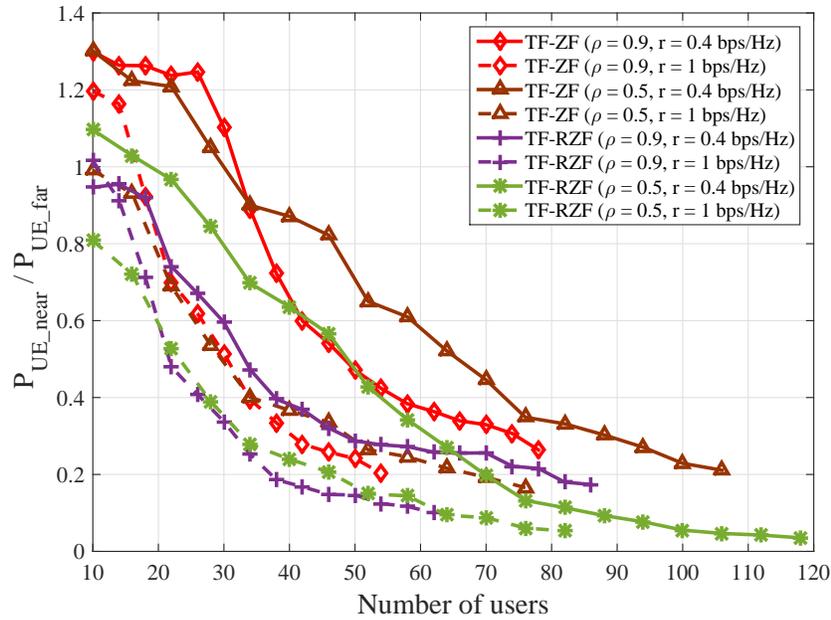}	
	\caption{The total transmit power for near UEs and far UEs ratio vs. the number of users.}
	\label{fig:ratio_pow_NoUE_3cell}
\end{figure}
Fig. \ref{fig:ratio_tau_NoUE_3cell} and Fig. \ref{fig:ratio_pow_NoUE_3cell} plot
the time-fraction ratio and power ratio, which are different from their counter parts in the above considered
single-cell and
two-cell cases. The number of near UEs during the time-fraction $\tau$ is half of
that during the time-fraction $1-\tau$ but the number of  cell-edge UEs during the former fraction
is double to that during the latter fraction. This fact dictates the allocation for both
time-fractions and powers.

\section{Conclusions}
We have considered the problem of maximizing the energy efficiency
in assuring the QoS for large numbers of users by multi-cell massive MIMO beamforming.
The antennas' spatial correlation, which is an important factor in assessing the actual
capacity of massive MIMO, has been incorporated in our consideration. To serve even larger numbers
of users within a time slot, techniques of  time-fraction-wise beamforming
have been proposed, including new path-following computational procedures for computational solution. The provided
simulations have demonstrated that $8$ $\times$  $8$ antenna array equipped massive MIMO is able to serve up to $160$ users
at required QoSs.

\section*{Appendix: fundamental inequalities}
By noting that function $f(x,y,t)=\frac{\ln (1+1/xy)}{t}$ is convex in $x>0, y>0, t>0$ \cite{Shetal17},  the following inequality for all $x>0$, $\bar{x}>0$, $y>0$, $\bar{y}>0$,  $t>0$, $\bar{t}>0$ holds true \cite{Tuybook}:
\begin{eqnarray}
\ds\frac{\ln(1+1/xy)}{t}&\geq& f(\bar{x}, \bar{y}, \bar{t})+\la \nabla f(\bar{x},\bar{y},\bar{t}), (x,y,t)-(\bar{x},\bar{y},\bar{t})\ra \nonumber\\
&=& \bar{a} - \bar{b} x - \bar{c} y - \bar{d} t,\label{ineq3}
\end{eqnarray}
and
\begin{eqnarray}
\ds \ln(1+1/xy) &\geq& a - b x - c y, \label{ineq4}
\end{eqnarray}
where $\nabla$ is the gradient operation and
\[
\begin{array}{c}
\bar{a} = 2\ds\frac{\ln(1+1/\bar{x}\bar{y})}{\bar{t}} + \frac{2}{\bar{t}(\bar{x}\bar{y}+1)} > 0,
\bar{b} = \frac{1}{(\bar{x}\bar{y}+1)\bar{x}\bar{t}} > 0,\\
\bar{c} = \ds\frac{1}{(\bar{x}\bar{y}+1)\bar{y}\bar{t}} > 0,
\bar{d} = \frac{\ln(1+1/\bar{x}\bar{y})}{\bar{t}^2} > 0,
\end{array}
\]
and
\[
a = \ln(1+1/\bar{x}\bar{y})+2/(\bar{x}\bar{y}+1) > 0,
b =1/(\bar{x}\bar{y}+1)\bar{x}  > 0,
c =1/(\bar{x}\bar{y}+1)\bar{y}  > 0.
\]
Replacing $x\rightarrow 1/x$ and $\bar{x}\rightarrow 1/\bar{x}$ in (\ref{ineq4}) leads to another inequality
\begin{eqnarray}
\ds \ln(1+x/y) &\geq& \tilde{a} - \tilde{b}/x - \tilde{c}y, \label{ineq4a}
\end{eqnarray}
for
\[
\tilde{a} = \ln(1+\bar{x}/\bar{y})+2\bar{x}/(\bar{x}+\bar{y}) > 0,
\tilde{b} =\bar{x}^2/(\bar{x}+\bar{y})  > 0,
\tilde{c} =\bar{x}/(\bar{x}+\bar{y})\bar{y}  > 0.
\]
Observing that function $f(z,t)={1}/{zt}$ is convex in $z>0, t>0$, we also have the following inequality
\begin{eqnarray}\label{ineq5}
\frac{1}{zt} &\geq& f(\bar{z},\bar{t})+\la \nabla f(\bar{z},\bar{t}), (z,t)-(\bar{z},\bar{t})\ra \nonumber \\
&=& 3\frac{1}{\bar{z} \bar{t}} - \left( \frac{z/\bar{z} + t/\bar{t}}{\bar{z} \bar{t}} \right), \quad \forall\
x>0,\ \bar{x}>0, t>0, \bar{t}>0.
\end{eqnarray}

\bibliographystyle{IEEEtran}
\bibliography{IEEEabrv,Report}
\end{document}